**Decoupling dynamics and crosslink stability in supramolecular hydrogels using associative exchange.**

*Pierre Le Bourdonnec, Charafeddine Ferkous, Leo Communal, Luca Cipelletti, and Rémi Merindol**

P. Le Bourdonnec, C. Ferkous, L. Communal, L. Cipelletti, and R. Merindol
Laboratoire Charles Coulomb, Université de Montpellier, CNRS, Montpellier France
E-mail: remi.merindol@umontpellier.fr

L. Cipelletti
Institut Universitaire de France, Paris, France.

Funding: This project has received financial support from the ANR through the JCJC programm Grant Number NR-20-CE06-0019 MIND and from the CNRS through the MITI interdisciplinary programs Grant Number 249744.

Abstract:

**The design of hydrogels that combine mechanical robustness with dynamic reconfigurability remains a fundamental challenge, as increasing crosslink dissociation rates compromise network integrity. This limitation is addressed through the incorporation of an associative crosslink exchange into DNA-based supramolecular hydrogels, enabling the decoupling of network relaxation behavior from crosslink stability. The hydrogels are constructed from enzyme-synthesized single-stranded DNA that self-assembles via hybridization between complementary domains. These crosslinks can reorganize through dissociative melting or associative strand displacement reaction, yielding networks with tunable relaxation timescales spanning over three orders of magnitude. Rheological measurements and thermodynamic modeling confirm that associative exchange facilitates efficient stress dissipation without diminishing rupture strength or thermal stability. In contrast, dissociative systems inherently trade increased dynamics with mechanical weakening. This decoupling is achieved through the implementation of a catalytic reorganization pathway governed by the composition of the sample, independently of crosslink strength. These findings establish the mechanism of**

**reorganization as a key design parameter for engineering adaptive soft materials that combine resilience and responsiveness.**

## 1. Introduction

Inspired by living tissues, hydrogels are water-swollen polymer networks that possess unique abilities to interact with biological systems. There is a growing interest in developing hydrogels for drug delivery, cartilage replacement, biomedical devices, and as artificial extracellular matrices.[1–3] Producing synthetic hydrogels with mechanical behaviors that closely mimic those of biological materials is of prime interest to facilitate biological interfacing. For implants and cartilage replacement, the focus is on designing strong and tough hydrogels capable of resisting fracture.[4,5] For cell culture, the emphasis is on engineering dynamic hydrogels that can replicate the cellular microenvironment.[6,7] We already know that altering the elastic modulus,[8–10] stress relaxation,[11,12] and stress stiffening properties[13–15] of the surrounding matrix has a profound impact on cell behavior. Surprisingly, biological tissues that consist of supramolecular networks, are both dynamic and tough.[16] These two properties are often mutually exclusive in synthetic systems, highlighting a gap in our understanding of how to decouple mechanical strength and dynamic behavior.[17]

Several key molecular parameters for the rational design of advanced hydrogels have already been identified. On the one hand, tailoring the network architecture primarily governs hydrogel elasticity.[18] Beyond linear mechanics, some of the most elegant structural designs incorporate slide-ring crosslinks, ideal network architectures, or interpenetrating networks to enhance ultimate stress and toughness.[19–21] On the other hand, dynamic crosslinking chemistries enable internal reorganization within the hydrogel network. Many strategies have been explored, including coordination bonds,[22,23] host–guest interactions, [24] dynamic covalent bonds, [25–27] and oligonucleotide self-assembly. [28–30] In such materials, the characteristic lifetime of the dynamic bonds dictates the network's relaxation behavior. Notably the network architecture and the lifetime of the crosslinks can be tuned independently to control the network's linear elasticity and relaxation time, respectively.[22,30] Yet, a fundamental trade-off persists between the strength and the reorganization speed of supramolecular hydrogels.[31] In dissociative networks, reorganization speed is controlled by the dissociation rate of the crosslinks, a concept summarized as "strong means slow".[32] Since dissociation is thermally driven, thermal agitation imposes an upper limit on the binding energy of crosslinks in such systems. Overcoming this limitation requires moving away from thermally driven dissociation and introducing a new design parameter: the reorganization mechanism.

The conflict between bond strength and dynamics in hydrogels mirrors a similar trade-off in bulk polymers. Thermosets are mechanically stable but cannot be recycled, whereas thermoplastics are reprocessable but suffer from creep and solubility in good solvents. Embedding catalytic reorganization mechanisms can address this challenge by enabling covalent bond rearrangement.[33] In particular, vitrimers, materials that undergo associative crosslink exchange, have gained significant atention.[34–36] Associative exchange involves an intermediate state where bond formation precedes bond dissociation, alowing melding and recycling of vitrimers without compromising on their strength and stability. However, due to the hydrolytic sensitivity of most vitrimer chemistries, associative crosslink exchange remains underexplored in hydrogels.[37] As a result, we lack a clear understanding of how the reorganization mechanism, either associative or dissociative, governs the macroscopic behavior of hydrogels. In this work, we propose a rational strategy to directly connect molecular reorganization mechanisms with the mechanical behavior of supramolecular hydrogels at the macroscale.

Beyond the storage of genetic information, DNA self-assembly offers one of the most powerful platforms for designing custom supramolecular materials. The hybridization of complementary DNA strands enables precise control over nanoscale architectures,[38,39] binding energies,[40–42] and even reorganization kinetics via strand displacement reactions.[43–45] Consequently, DNA has emerged as a platform of choice for engineering dynamic, responsive, and biocompatible hydrogels.[46–48] DNA crosslinks are inherently dissociative: when sufficient thermal energy is applied, the hydrogen bonds between complementary strands break, leading to crosslink rupture.[49] Yet, associative crosslink exchange can be programmed using strand displacement reactions.[29,50] This makes DNA hydrogels an ideal system to systematically investigate how reorganization mechanisms influence mechanical behavior. So far, however, this could not be tested in macroscale samples: systematic rheological studies of macroscale DNA hydrogels are rare, largely due to the high cost of synthetic oligomers. We overcome this limitation using rolling circle amplification (RCA), an isothermal enzymatic process that produces milligram quantities of sequence-controlled DNA and facilitates hydrogel assembly.[47,51]

We present a strand displacement-based reorganization mechanism that implements associative crosslink exchange in macroscale all-DNA hydrogels produced by RCA. Crucially, this design enables a direct, systematic comparison between dissociative and associative mechanisms since it allows for switching between dissociative and associative exchange without altering the network structure or crosslink stability. Despite this potential, such a comparison has not yet been achieved in macroscale samples. Systematic. We correlate the thermodynamics of these molecular mechanisms with the macroscale mechanical behavior of the hydrogels, both in the linear viscoelastic regime and the non-linear regime, where bond strength plays a dominant role. Our systematic approach, combining controlled supramolecular design with rigorous macroscale mechanical characterization, highlights the critical role of reorganization mechanisms as a design parameter for next-generation hydrogels.

## 2. Assembly of Dynamic DNA Hydrogels

The DNA synthesis and hydrogel assembly are presented in **Figure 1**. The enzymatic synthesis (RCA) uses a small circular template and nucleoside triphosphates as inputs and yields long single-stranded DNA consisting of multiple copies of the template linked one after another (see also Figure S1 for details). Macroscale hydrogels are formed by mixing RCA products with complementary domains (e.g., α and α*). After heating above 95 °C to melt all DNA duplexes and cooling down to room temperature, we obtain an homogeneous supramolecular network held together by DNA duplexes (see also Figure S2 for details).[47] Such a supramolecular DNA hydrogel is dissociative (Figure 1B). When provided with sufficient thermal energy, the complementary domains can spontaneously melt and rehybridize, allowing network reorganization. Such dissociative reorganization is common in supramolecular networks.[17,22,24,49]

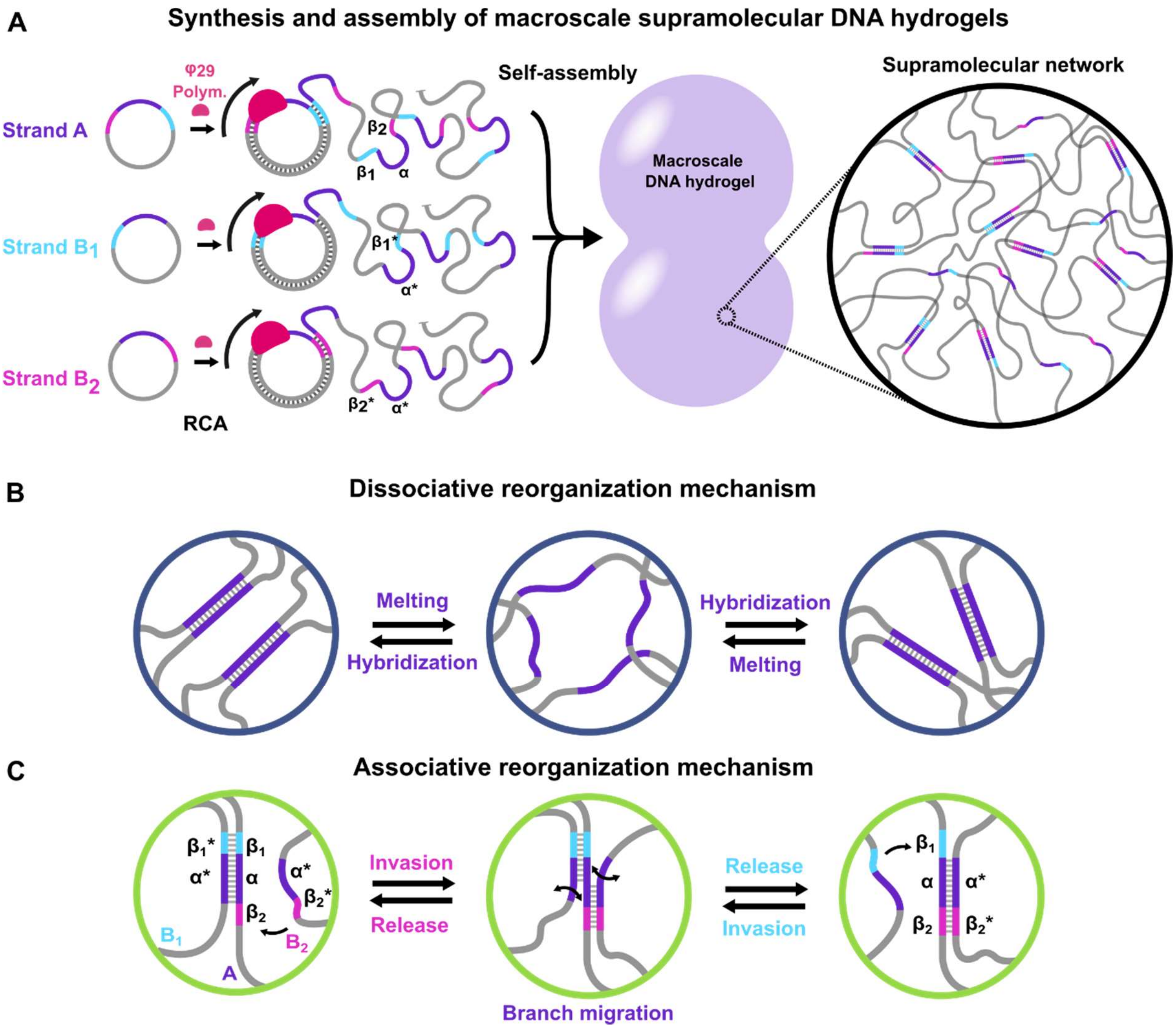

**Figure 1** A) Schematic representation of the DNA strands synthesized by rolling circle amplification (RCA) and the network architecture of the macroscale supramolecular DNA hydrogels resulting from their assembly. B, C) Key mechanisms controlling crosslink reorganization. B) Dissociative reorganization driven by the melting and rehybridization of a crosslinking duplex. C) Associative exchange driven by the invasion, branch migration, and release of a crosslinking duplex (i.e., a strand displacement reaction).

Here, we further engineer an associative exchange using a strand displacement reaction programmed into the repeat sequences of three RCA products: a pivot strand A and two exchange strands $B_1$ and $B_2$ (**Table 1**). The repeat sequence of the pivot strand consists of a crosslink domain α, flanked by two toeholds domains $\beta_1$ and $\beta_2$. The repeat sequences of the exchange strands consist of a crosslink-complementary domain $\alpha^*$ flanked by one of the toehold-complementary domains, either $\beta_1^*$ or $\beta_2^*$. For each strand, the functional domains are separated on the RCA product by an adenine spacer ($A_{30}$), which provides flexibility. Upon mixing, the crosslink domains (α and $\alpha^*$) of the exchange and pivot strands hybridize, forming the 3D network. These sequences also enable associative exchange through the strand displacement reaction presented in Figure 1C. Starting from the left, a crosslink formed by a pivot-exchange duplex A/$B_1$ leaves accessible the toehold domain $\beta_2$, while the domains $\alpha/\alpha^*$ and $\beta_1/\beta_1^*$ are hybridized. The strand displacement reaction starts when an exchange strand $B_2$ hybridize to the free toehold $\beta_2$ to invade the duplex A/$B_1$. In the middle panel, $B_2$ progressively replaces $B_1$ via branch migration. On the right, $B_2$ has replaced $B_1$ at the α domain, the toehold domain $\beta_1/\beta_1^*$ melts, and $B_1$ is released. This strand displacement reaction is fully reversible: starting from the right panel, strand $B_1$ can invade the duplex A/$B_2$ to replace and release $B_2$. The toehold domains $\beta_1$ and $\beta_2$ are short enough (6 bases) to spontaneously hybridize and melt at room temperature, allowing the exchange strands $B_1$ and $B_2$ to switch from one crosslink to another.

**Table 1.** Name, repeat sequences and functional domains of the RCA products.

| RCA product | Repeat sequence 5'→3' a) | Functional domains |
|---|---|---|
| Pivot strand A | [CACCGA GCACAGCGTCGAGG AGCACC $A_{30}]_n$ | $[\beta_1\ \alpha\ \beta_2\ A_{30}]_n$ |
| Exchange strand $B_1$ | [CCTCGACGCTGTGC TCGGTG ACCTATACGT $A_{30}]_n$ | $[\alpha^*\ \beta_1^*\ \delta_1\ A_{30}]_n$ |
| Exchange strand $B_2$ | [TGTTAGTAGT GGTGCT CCTCGACGCTGTGC $A_{30}]_n$ | $[\delta_2\ \beta_2^*\ \alpha^*\ A_{30}]_n$ |

a) We use a space to separate and identify each each functional domain in the repeat sequence.

The sequences are designed so that the free energy of both duplexes A/$B_1$ and A/$B_2$ are equal ($\Delta G_{A/B1} = \Delta G_{A/B2} = \Delta G_{S0 \rightarrow S2} = -12$ kJ/mol at 25 °C, Table S1).[52] There is therefore no energetic preference for A to bind to $B_1$ or $B_2$, which prevents the exchange from being terminated due to saturation with a preferred exchange strand. This is an associative reorganization mechanism, since crosslink connectivity increases during the reconfiguration process. In contrast, reorganization driven by duplex melting is dissociative, as the crosslink connectivity decreases during the reconfiguration process.

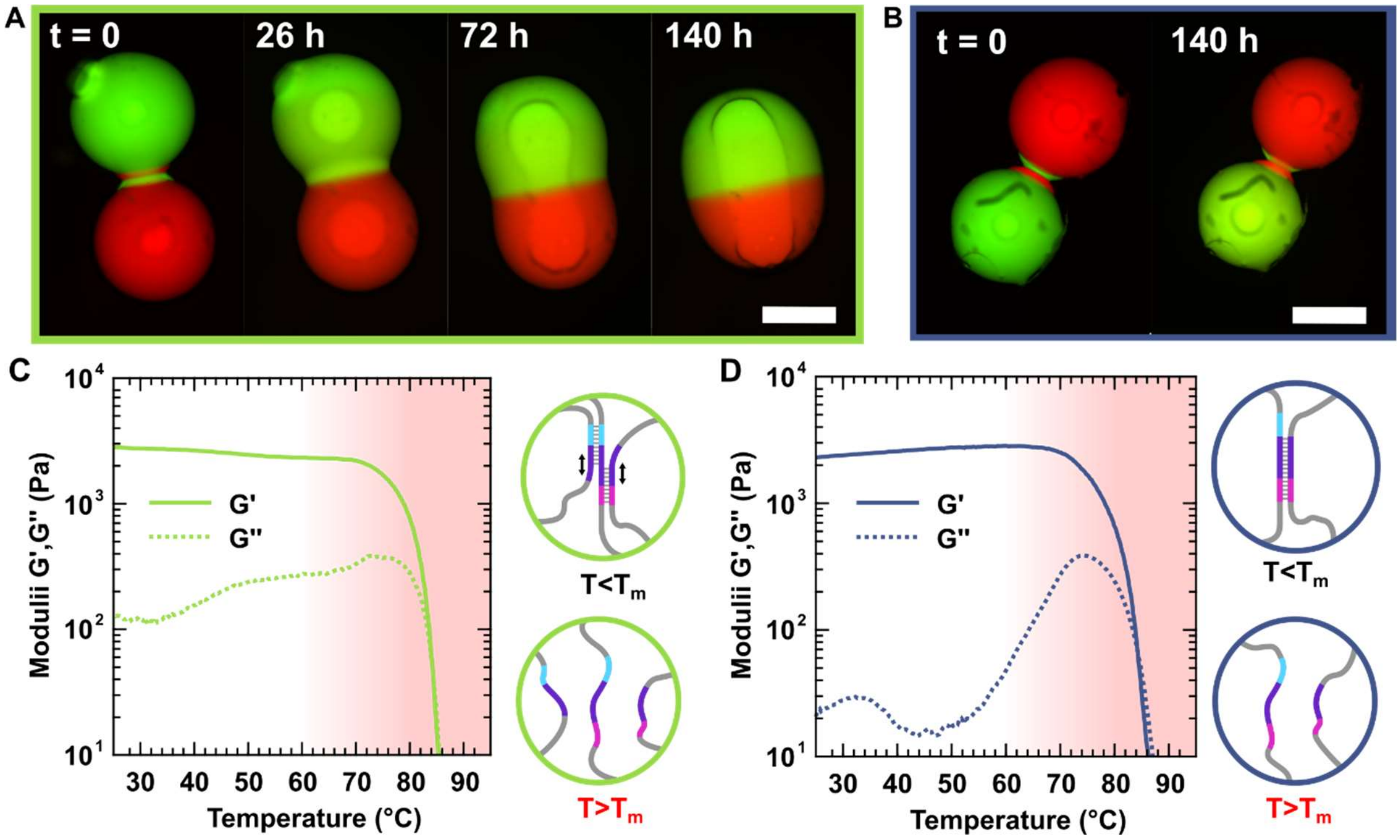

**Figure 2** Key features of associative and dissociative DNA hydrogels. A) Wide-field microscopy images showing associative DNA hydrogels melding at 30 °C. B) Under the same conditions, dissociative hydrogels cannot reorganize and meld. The hydrogels are immersed in mineral oil to force contact via capillary forces and to prevent water evaporation. Here the DNA strands are fluorescently labeled during synthesis by replacing some of the thymine bases with $Atto_{488}$- or $Atto_{565}$-modified bases (See Note S1). Scale bar = 0.5 mm. C,D) Evolution of the storage and loss moduli with temperature for associative (C) and dissociative (D) DNA hydrogels. The schemes to the left of each graph represent the predominant crosslinking state below and above $T_m$, respectively. All hydrogels are composed of the same DNA strands and differ only in the relative proportion of strands A and B. Associative hydrogels contain a two-fold excess of strand B compared to strand A, whereas dissociative hydrogels have a slight deficit of strand B relative to A. Oscillatory rheology data are collected on 2 wt% DNA hydrogels at 1 rad·s$^{-1}$ and 5% strain.

The nature of the reorganization mechanism, associative or dissociative, significantly influences the macroscopic behavior of DNA hydrogels at room temperature. The characteristic behaviors of DNA hydrogels with and without associative crosslink exchange are presented in **Figure 2**. The associative crosslink exchange is catalyzed by the presence of free toeholds $\beta_1^*$ or $\beta_2^*$ : therefore it requires an excess of sequences $B_1$ and $B_2$ compared to sequence A in order to proceed. To facilitate comparison, all DNA hydrogels presented in this work are made from the same RCA products A, $B_1$, and $B_2$. The sequences $B_1$ and $B_2$ play symmetrical roles and are always present in equal amounts. For simplicity, we use B to refer collectively to sequence $B_1$ and $B_2$ (i.e., [B] = [$B_1$] + [$B_2$]) and $\beta^*$ to refer collectively to $\beta_1^*$and $\beta_2^*$. The only variable controlling the reorganization mechanism is the ratio of sequence A to B. When B sequences are in excess, the hydrogels are associative, because they can reorganize via strand displacement reactions. If A strands are in excess, there are no free toeholds $\beta^*$ available to initiate a strand displacement reaction, and the hydrogel is dissociative. This compositional difference strongly impacts hydrogel dynamics at room temperature. Associative hydrogels can meld at 30 °C (Figure 2A), whereas dissociative hydrogels do not (Figure 2B) : the associative crosslink exchange increases the reorganization rate of DNA networks and enables macroscopic reshaping at room temperature.

Associative and dissociative DNA hydrogels also display distinct viscoelastic signatures. We systematically measure the loss and storage shear moduli of associative and dissociative hydrogels as a function of temperature $T$ (Figure 2C-D). In these oscillatory shear experiments, the storage and loss moduli represent the elastic and viscous contributions to the material response, respectively. At low $T$, associative hydrogels are more dissipative than dissociative ones : they exhibit a tenfold higher loss modulus at room temperature. In contrast, because both hydrogels have similar network architectures, their storage moduli are comparable. For associative hydrogels, the changes in loss and storage moduli are small and gradual up to 75 °C (Figure 2C). We observe a slight increase in the loss modulus and a decrease in the storage modulus, which results from an increase in the exchange rate with temperature. The small amplitude of these variations demonstrates that such hydrogels are remarkably stable over a broad temperature range, while showing significant reorganization even at low temperatures, as demonstrated by their ability to meld (Figure 2A). Above 75 °C, both the storage and loss moduli drop rapidly. This marks the melting transition of the crosslinks, a dissociative process that progressively replaces the associative crosslink at high $T$ : the DNA duplexes (i.e., crosslinks) shift from a closed to an open state, resulting in loss of network integrity.

For dissociative hydrogels, both the loss and storage moduli remain nearly constant up to 60 °C. At these temperatures, the crosslinks are stable but not dynamic, the hydrogel does not reorganize within a reasonable timescale, as shown by the absence of melding (Figure 2B). Between 60 °C and 75 °C, we observe a sharp increase in the loss modulus: thermal energy enables dissociative crosslink reorganization, increasing the viscous response of the sample. The proportion of dissociated crosslinks increases with temperature. Below ~75 °C, this proportion is small enough for its effect on the hydrogel's elastic response to remain negligible.

However, as for associative hydrogels, above 75 °C, the network loses its structural integrity as the duplexes melt. In dissociative reorganization, the temperature range in which dynamic behavior and structural stability coexist is relatively narrow. For example, in dissociative hydrogels, the fraction of dissipated energy, expressed as *tan(δ) = G''/G'*, decreases from approximately 0.25 at 75 °C to 0.05 at 65 °C. In comparison, the associative hydrogels shows a similar *tan(δ)* of ~0.25 at 75 °C but reaches 0.05 only at 25 °C. Thus, the presence of an associative crosslink exchange mechanism increases by a factor of five the temperature range over which a solid hydrogel exhibits significant dissipation. This is a key signature of the associative exchange mechanism, which enables decoupling of reorganization dynamics at room temperature from the thermal stability of the crosslinks.

### 3. Controlling the reorganization dynamics using strand displacement reaction.

The main control parameter of the associative exchange mechanism is the concentration of active catalytic toeholds β*. As discussed earlier, the strand displacement reaction requires a free exchange strand to proceed. To compare samples of different compositions, we define the stoichiometric ratio *R*, which gives the proportion of active toeholds relative to the number of crosslinks (**Equation 1**):

$$R = \frac{[B]-[A]}{[X_{AB}]} \qquad \text{Equation 1}$$

Where [A] and [B] are the total concentrations of the repeat sequences of strands A and B, respectively, and $[X_{AB}]$ is the total concentration of crosslinks (either $A/B_1$ or $A/B_1$). Since the formation of a crosslink requires one α and one α* domain, the concentration of crosslinks is given by $[X_{AB}]$ = min([A], [B]). The scheme in **Figure 3A** summarizes the effect of the ratio *R* on the associative exchange mechanism. For $R \leq 0$, all toeholds β* on the B strands are involved in A/B duplexes, so no active toeholds remain to initiate strand displacement. Note that, the excess of toehold domains on A strands cannot hybidize to A/B crosslinks to initiate strand displacement. Therefore, for $R \leq 0$, crosslinks cannot reorganize via an associative exchange mechanism. In contrast, for $R > 0$, the hydrogel contains an excess of exchange strands, meaning that active toeholds β* are available for the associative exchange to occur. By design, *R* is directly proportional to the number of available toeholds capable of initiating strand displacement ; therefore, the reorganization rate via associative exchange scales with *R*.

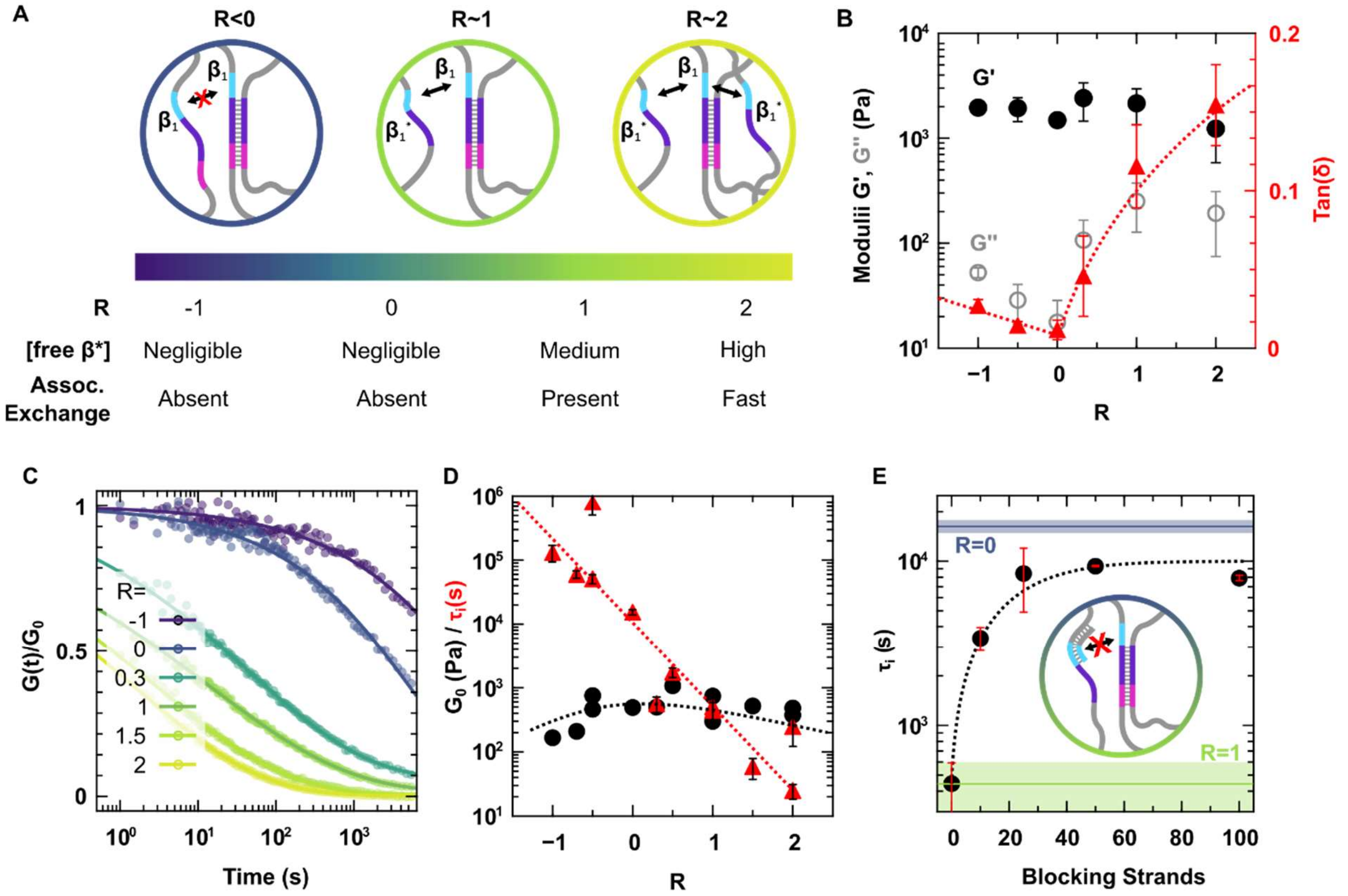

**Figure 3**: Controlling the reorganization dynamics with associative crosslink exchange. A) Schematic representation of the effect of sample composition in strands A and B on the availability of catalytic toeholds β* and the resulting reorganization mechanism. B) Evolution of the elastic (G′, black closed circles) and viscous (*G″*, grey open circles) moduli, and of the loss factor *tan(δ)* (red) of DNA hydrogels as a function of the stoichiometric ratio *R*, as defined in Equation 1. Data were acquired on 2 wt% hydrogels at 37 °C, 1 rad·s$^{-1}$, and 5% deformation. Error bars represent the standard deviation across two to five different samples. C) Stress relaxation experiments for hydrogels of with different *R* (all data: 5% strain, 37 °C). Symbols represent experimental stress measurements and solid lines are stretched exponential fits (see Equation 2). D) Initial relaxation modulus $G_0$ (black circles) and integral relaxation time $\tau_i$ (red triangles) obtained by fitting the stress relaxation data using a stretched exponential decay. Error bars on $G_0$ are too small to be visible. E) Effect of adding blocking strands that sequester toeholds (see inset) on the integral relaxation time of an associative hydrogel (*R* = 1). For comparison, green and blue lines represent the integral relaxation times of *R* = 1 and *R* = 0 hydrogels, respectively, without blocking strands; the shaded areas indicate the corresponding error margins. In panels B, D, and E, dotted lines are included as guides to the eye.

We systematically measured the rheological behavior of DNA hydrogels as a function of their composition $R$. First, we examined the storage (G′) and loss (G″) moduli of the hydrogels in the linear regime (Figure 3B). The loss modulus increases with $R$, i.e., as strand displacement becomes more prevalent, while the storage modulus remains relatively constant. This effect is more clearly observed in the loss factor, *tan(δ) ′*, which characterizes the relative contribution of dissipation to the mechanical response of the sample. For $R \leq 0$, *tan(δ)* is low and nearly constant. As expected, it steadily increases with $R$ for $R > 0$. These measurements confirm that the proposed associative reorganization mechanism is stoichiometry-controlled. To quantitatively assess the timescale of reorganization, we performed stress relaxation experiments (Figure 3C). We fitted the results with a stretched exponential decay, as defined in **Equation 2**.

$$\sigma(t) = \gamma G_0 e^{-(\frac{t}{\tau_r})^{\beta}} + s \qquad \text{Equation 2}$$

where $G_0$ is the initial shear modulus, $\gamma$ the imposed step strain, $\tau_r$ the 1/e relaxation time, and $\beta$ the stretching exponent. s is a small baseline correction applied to dynamic samples only, to avoid fitting artifacts arising from experimental uncertainty near the $\sigma \approx 0$ baseline. A simple exponential decay ($\beta = 1$) corresponds to a single relaxation time, but the stretching exponents measured in our samples ($0.2 \leq \beta \leq 0.5$) indicates a broad distribution of relaxation times. This distribution is better captured by the integral relaxation time ($\tau_i$) rather than $\tau_r$, although both show similar trends (see also Note S4 and Figure S4 for details). In Figure 3D, we report both $\tau_i$ and the initial relaxation modulus ($G_0$) as a function of $R$. Consistent with the oscillatory shear data, the initial relaxation modulus $G_0$ remains relatively constant across compositions, while the relaxation time decreases from $\tau_i > 10^5$ s for dissociative hydrogels ($R \leq 0$) to $\tau_i < 200$ s for associative hydrogels ($R > 1$). Thus, the catalytic control of the associative exchange mechanism provides a convenient means of tuning relaxation times across more than three orders of magnitude. The broad distribution of relaxation times may seem surprising, as dissociative networks typically exhibit narrower distributions. However, the network consists of long DNA strands with multiple sticky domains, so relaxation proceeds via sticky reptation, where complete relaxation requires the sequential breaking and reforming of multiple crosslinking points.[53,54] The presence of associative crosslink exchange, which prevents some of the topological relaxation permitted in dissociative networks, further amplifies this effect. If needed, such structural contributions can be mitigated by modifying the network topology or by decreasing the length of the RCA products.[29]

We have shown the possibility of controlling the reorganization of DNA hydrogels by changing their composition. Importantly, one can also control the reorganization mechanism of an already formed sample. The control of strand displacement reactions using short oligomers that hide or reveal toeholds is well established.[43] Here, we leverage this approach by designing "blocking strands" that hybridize to an addressing domain δ (either $\delta_1$ for $B_1$ or $\delta_2$ for $B_2$) and hide the toeholds β* to suppress associative exchange. The addressing domain is necessary to stabilize the hybridization of the blocking strands to the network. As mentioned earlier, without stabilization, a six-base-long strands complementary to the toeholds would continuously melt and rehybridize at room temperature without blocking the associative exchange. As shown in Figure 3E, adding 10 mol% of blocking strands (relative to the total number of β* toeholds) increases the relaxation time of an $R = 1$ hydrogel from 200 $s$ to 2000 $s$. Further increasing the concentration of blocking strands to 25–50% of toehold concentration increases the relaxation time up to 9000 $s$, comparable to that of a dissociative hydrogel with $R = 0$. Note that we expect no additional effect from increasing the concentration of blocking strands beyond 50% of toeholds. Indeed, for $R = 1$, 50% of the toeholds β* are already engaged in crosslinks and thus inactive; therefore, one expects 50% of blocking strands to be sufficient to block 100% of the remaining catalytically active toeholds. Interestingly, we observe a plateau in relaxation time at 25%, earlier than expected, possibly because blocking a sufficiently large fraction of the available toeholds already introduces enough constraints in the relaxation process to significantly slow it. While the origin of this premature saturation remains to be fully clarified, these experiments demonstrate that blocking strands are highly effective in controlling in-situ the reorganization speed of already formed dynamic hydrogels.

## 4. Thermodynamic understanding of the reorganization mechanism.

We expect the reorganization speed of dynamic hydrogels to increase with temperature, as higher temperatures generally accelerate chemical reaction rates. Investigating how reorganization dynamics vary with temperature can further offer valuable insights into the underlying thermodynamic landscape governing the crosslink exchange reactions. We performed frequency sweeps on DNA hydrogels with stoichiometric ratios varying from $R = -1$ to $R = 2$, at multiple temperatures ranging from 15 °C to 85 °C. We constructed master curves for $G'$ and $G''$ by applying the Time–Temperature Superposition (TTS) principle (details in Note S5).[55] The characteristic master curves for associative ($R = 1$) and dissociative ($R = -0.5$) hydrogels are shown in **Figure 4A**. Overall, both master curves exhibit Maxwell-like features, including terminal relaxation at low frequency, a maximum in $G''$ near the $G'/G''$ crossover, and a plateau at high frequencies. However, the overall shape of the master curves deviates from that of an ideal Maxwell material, as both dynamic and static hydrogels reach the peak of dissipation at frequencies higher than the *G′/G″* crossover. This is typical of materials with a broad distribution of relaxation times, as previously identified in stress relaxation experiments (Figure 3C,D).[54] The dissociative hydrogels display a well-defined dissipation peak with a single maximum, suggesting that a single chemical process governs relaxation. In contrast, the dissipation peak of associative hydrogels exhibits a shoulder, suggesting the presence of multiple relaxation mechanism.[22] We attribute this feature to the transition from associative exchange at low temperatures to dissociative exchange at high temperatures.

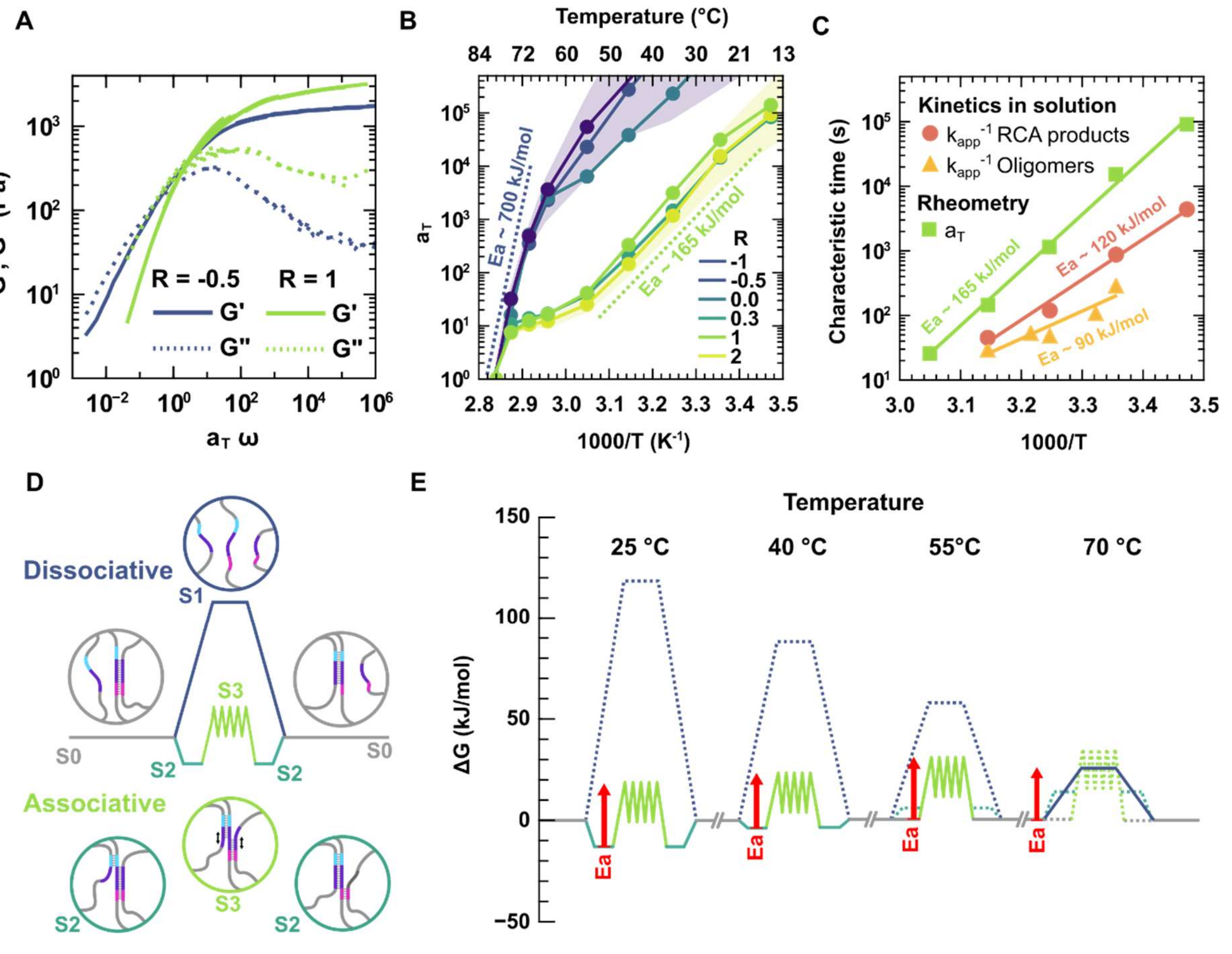

**Figure 4**: Rationalizing the temperature dependence of associative and dissociative exchange mechanisms. A) Rheological master curves of dissociative (*R* = -0.5) and associative (*R* = 1) DNA hydrogels obtained by time-temperature superposition of frequency sweeps performed between 15 °C and 85 °C, using the shift factors $a_T$ shown in (B). The reference temperature is 85 °C, where both hydrogels reorganize via dissociative mechanisms and exhibit similar dynamics. B) Arrhenius plot showing the evolution of the shift factor $a_T$ with *1000/T* for DNA hydrogels with different compositions. The shaded areas in blue and green indicate the error intervals for *R* = -0.5 and *R* = 1, respectively (see Figure S5 for details). The slopes of the curves are proportional to the activation energy of the dominant reorganization mechanism at a given

temperature. C) Evolution of the characteristic time of associative exchange with *1000/T*, measured as $k_{app}^{-1}$ for model reactions using short oligomers and long RCA products in solution (yellow triangles and red circles, respectively), plotted alongside the corresponding $a_T$ values obtained by TTS for macroscopic hydrogels (green squares). See Note S5 and Figure S6 for details. D) Schematic representation of the free energy profiles for dissociative (top, blue) and associative (bottom, green) crosslink exchange mechanisms. The curves represent the evolution of the free energy ($\Delta G$) of intermediate conformations for both mechanisms. The $S_3$ state is shown as a sawtooth profile, since during branch migration, each nucleobase step is associated with a variation in $\Delta G$.[56] E) Quantitative evolution of the energetic profiles of exchange with temperature (see Note S6 and Figure S7 for details). The solid lines represent the exchange pathway (associative or dissociative) that minimizes the activation energy ($E_a$) of the exchange (shown in red) at a given temperature. The dotted lines indicate the alternative, less favorable pathways, shown for comparison.

The shift factors $a_T$ obtained from TTS reflect the underlying reorganization mechanism. The values of $a_T$ for hydrogels with stoichiometric ratios from $R$ = -1 to $R$ = 2 are presented in Figure 4B. Strikingly, the evolution of $a_T$ with temperature depends only on the type of reorganization mechanism, not on the exact value of $R$. Hydrogels with associative mechanisms ($R > 0$) collapse onto a single $a_T$ curve, while dissociative hydrogels ($R \leq 0$) collapse onto a different curve. For materials that reorganize via reaction-controlled processes, such as DNA melting or strand displacement, $a_T$ is proportional to the reaction rate constant $k$. Therefore, $a_T$ follows an Arrhenius dependence on temperature, as described in **Equation 3** :

$$a_T \propto k = A \cdot e^{\left(-\frac{E_a}{R_g \cdot T}\right)} \qquad \text{Equation 3}$$

where $E_a$ is the activation energy of the reaction, $R_g$ the ideal gas constant, $T$ the absolute temperature, and $A$ the Arrhenius prefactor of the reaction. By plotting $a_t$ as a function of *1000/T* (i.e., in an Arrhenius plot), we can identify the temperature ranges where the reorganization is dominated by a single chemical process (Figure 4B). At low temperatures, between 15 °C and 55 °C, all associative hydrogels ($R > 0$) exhibit similar slopes, corresponding to an activation energy of 165 ± 20 kJ/mol. We attribute this energy barrier to the rate-limiting step of the strand displacement reactions. Dissociative hydrogels, on the other hand, display a significantly higher activation energy in the range of ~700 kJ/mol, measurable between 50 °C and 85 °C. At lower temperatures, the evolution of the modulus with $T$ is too small to allow reliable measurement (the error on $a_T$ becomes very large), while at higher temperatures the network loses its structural integrity and the moduli become too low to measure.

Interestingly, we find that associative hydrogels show a similar slope to dissociative ones between 75 °C and 85 °C, indicating that the dissociative pathway dominates reorganization at high temperatures. Surprisingly, we also observe a flattening of the Arrhenius slope between 60 °C and 75 °C, suggesting a shift in the rate-limiting step controlling the associative crosslink exchange. These differences in activation energy can be rationalized based on the thermodynamic profile of the respective exchange mechanisms.

Discussing activation energies from a thermodynamic perspective requires some care. It is generally accepted that for multistep reactions, the activation energy ($E_a$) is determined by the free energy barrier of the rate-limiting step (i.e., $E_a \approx \Delta G^*$). The lower the $\Delta G^*$, the faster the reaction proceeds at a given temperature. However, since $\Delta G^*$ itself varies with temperature, the slope of the curves in an Arrhenius plot is not directly proportional to $\Delta G^*$. By decomposing the free energy variation $\Delta G^*$ into its temperature-independent thermodynamic components, enthalpy ($\Delta H^*$) and entropy ($\Delta S^*$), we obtain the Eyring–Polanyi equation (**Equation 4**). [57,58]

$$a_T \propto A \cdot e^{\left(-\frac{\Delta G^*}{R_g \cdot T}\right)} = A \cdot e^{\left(-\frac{\Delta H^* - T\Delta S^*}{R_g \cdot T}\right)} = A \cdot e^{\left(-\frac{\Delta S^*}{R_g}\right)} \cdot e^{\left(-\frac{\Delta H^*}{R_g \cdot T}\right)} \quad \text{Equation 4}$$

From Equation 4, we see that the slope of the Arrhenius plot corresponds to the enthalpy change $\Delta H^*$ of the rate-limiting step. Nevertheless, it is the free energy landscape that ultimately determines the absolute rate of a reorganization mechanism, and therefore which mechanism dominates at a given temperature.

The thermodynamic parameters controlling DNA hybridization and strand displacement reactions are well established. Leveraging the conceptual framework of the energy landscape elucidated by Winfree et al.[56] and UNAFold simulations,[52,59] we present a quantitative diagram of the reaction pathways, in free energy, for the associative and dissociative crosslink exchanges across a range of temperatures (Figure 4E, also Note S6 and Table S1 for details). As expected, at room temperature, the activation energy for dissociative exchange (blue line) is much higher than that of associative exchange (green line). Yet, since the configurational entropy of the dissociated state ($S_1$) is higher than that of the associated 3-strand complex ($S_3$), the activation energy for the dissociative exchange path decreases more rapidly with temperature than that of the associative exchange path, due to the contribution of the $-T\Delta S$ term in $E_a$. The activation energies of the two paths become comparable around 60 °C. The profiles also predict an intermediate regime, between 45 and 60 °C, where toehold hybridization is unfavorable ($\Delta G > 0$), but where the associative exchange still minimizes the activation energy.

In this regime, illustrated by the diagram at 55 °C in Figure 4E, the exchanges proceed via direct invasion of the duplex without a stable $S_2$ state. This intermediate regime has a distinct $\Delta H^*$ (Figure S7) and explains the variation in activation energy observed in the TTS analysis between 60 and 75 °C. A noteworthy observation is the consistent high temperature shift of the experimentally measured transitions in TTS relative to the predictions derived from our thermodynamic profiles. This discrepancy is primarily attributable to the fact that the thermodynamic parameters used to construct the energy landscape are taken from studies on short oligomers in dilute solution, thereby omitting complex contributions arising from network effects.[56,60] Nevertheless, such detailed reaction profiles offer a robust framework for the qualitative identification of reaction pathways and the elucidation of rate-limiting steps controlling molecular reorganization at different temperatures.

We also compare quantitatively the model predictions of the enthalpy variation ($\Delta H^*$) for each rate-limiting step (Table S1) with the slopes measured in the Arrhenius plot. On the one hand, for the associative exchange, the predicted activation energy, $\Delta H_{S2\rightarrow S3}$ = 88 kJ/mol, is lower than the value obtained from TTS, $E_a$ = 165 ± 20 kJ/mol. Here again, we expect that topological constraints imposed by the network on the dangling ends of the crosslinks affect the strand displacement reaction thermodynamics. To test this hypothesis, we designed a model strand displacement reaction consisting of small DNA oligomers that mimic the associative crosslink exchange found in our hydrogel and that can be monitored by fluorimetry (See Note S5 and Figure S6 for details). By measuring the apparent exchange rate $k_{app}$ of this model reaction between 15 and 55 °C, we plot the evolution of the characteristic times (taken as $k_{app}^{-1}$) in an Arrhenius plot (Figure 4C, triangles). We find an activation energy of 90 ± 15 kJ/mol, in excellent agreement with the thermodynamically predicted $\Delta H_{S2\rightarrow S3}$= 88 kJ/mol. For long RCA products in dilute conditions, we find an activation energy of 120 ± 10 kJ/mol, intermediate between those of the 3D networks and the oligomers. Note that in dilute conditions, RCA products are expected to form small clusters with fewer topological constraints than in 3D networks, but more than for isolated oligomers. This confirms that the enthalpy variation of the associative crosslink exchange is significantly influenced by topological constraints on the dangling ends. On the other hand, the values predicted for the dissociative reaction pathway ($\Delta H_{S0\rightarrow S1}$= 735 kJ/mol) match those found for dissociative hydrogels. Overall, simple reaction profiles such as the one developed here, though they omit network contributions, are in good semiquantitative agreement with the data and represent a valuable tool to identify the key mechanisms controlling network reorganization at equilibrium. Next, we demonstrate their use in rationalizing the rupture of our DNA hydrogels under mechanical stress.

### *5.* Decoupling dynamics and rupture with associative exchange mechanism.

We have shown that the presence of an associative exchange enables the decoupling of hydrogel dynamics from thermal stability. Here, we take a step further by examining how exchange mechanisms influence the rupture of the network. Ultimately, the rupture strength of a supramolecular hydrogel is determined by the intrinsic strength of its crosslinks, although the overall network architecture also plays a significant role.[4] For dissociative hydrogels, the rupture and relaxation pathways are identical, both involving the opening of crosslinks. In contrast, associative hydrogels offer two possible pathways (**Figure. 5A**). On one hand, the associative crosslink exchange is active at rest under ambient conditions due to its low activation energy, but its rate is limited by the concentration of active catalytic domains (free β*). This pathway governs the relaxation dynamics of unperturbed samples, as previously discussed. On the other hand, the dissociative pathway is inactive at room temperature due to its high activation energy, but can be mechanically activated under stress. This dissociative pathway governs the hydrogel's rupture strength at high strain rates, since its rate is not limited by the presence of catalytic sites. In startup shear experiments, both pristine dissociative and associative hydrogels exhibit similar rupture strengths (Figure. 5B,C), indicating that their rupture proceeds via similar dissociative mechanisms. However, only the associative hydrogel can self-heal, within 24 hours, thanks to its low-activation-energy reorganization mechanism. Thus, the associative exchange mechanism enables self-healing without compromising the mechanical strength of the material.

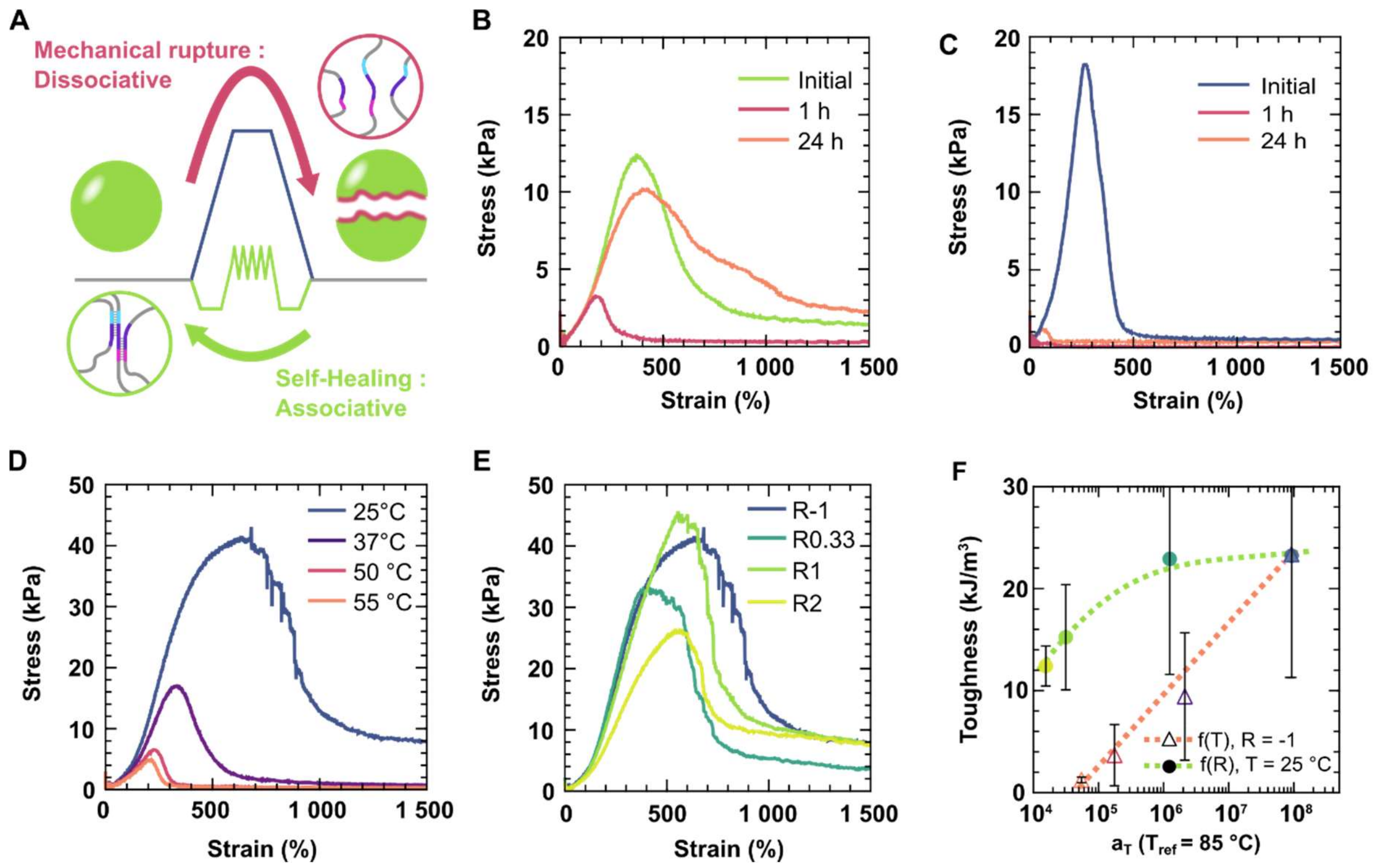

Figure 5: Rupture and self-healing of associative and dissociative DNA hydrogels. A) Schematic representation of the two different pathways for rupture and self-healing of associative DNA hydrogels. Rapid mechanical rupture involves the dissociation of the DNA crosslinks, which have a high activation energy at room temperature. In contrast, self-healing is slower but relies on a strand displacement reaction that has a low activation energy. B,C) Start-up shear experiments on DNA hydrogels with and without associative exchange, respectively, as prepared (initial), and after 1 hour or 1 day at 37 °C after rupture. D) Start-up shear experiments for dissociative DNA hydrogels performed at various temperatures. The sample weakens with increasing temperature. E) Start-up shear experiments for associative DNA hydrogels with increasing stoichiometric ratio *R*. The sample strength is almost independent of *R*. F) Systematic comparison of toughness (measured as the area under the stress–strain peak) as a function of reorganization dynamics, quantified by the shift factor $a_T$ obtained from time–temperature superposition (TTS). Error bars correspond to the standard deviation from measurements on three different hydrogels.

While both dissociative and associative reorganization pathways control hydrogel relaxation, they affect toughness differently. In dissociative mechanisms, the hydrogel's relaxation time is dictated by the dissociation rate of crosslinks, which can be modulated by temperature.[32] Increasing temperature lowers the activation energy ($E_a$) of the dissociative pathway (Figure 4E), thereby accelerating relaxation. In contrast, in associative mechanisms, the characteristic relaxation time is governed by the concentration of catalytically active domains, quantified by the stoichiometric ratio $R$ (Figure 3D). To compare how each mechanism affects hydrogel strength, we performe startup shear tests while varying either $T$ or $R$, keeping the shear rate constant at 1 $s^{-1}$. Figure 5D shows the rupture behavior of dissociative hydrogels ($R$ = -0.5) as the temperature increase from 25 °C to 75 °C. While the slope at the origin remains unchanged, consistent with the nearly constant $G'$ over this temperature range (Figure 2D), the rupture strength (i.e., the maximum stress reached during startup shear) decreases rapidly with temperature. This behavior contrasts with that of associative hydrogels tested at constant temperature (T = 25 °C) while varying composition ($R$ = [-0.5; 2], Figure. 5E). In this case, both the slope at the origin and the rupture strength remain relatively constant. Only the fastest-reorganizing samples ($R$ = 2) exhibits a slight decrease in rupture stress, likely due to rapid rearrangement enabling partial stress relaxation prior to failure. To quantify the toughness, defined as the energy stored before rupture, we calculate the area under the stress–strain curves and compare it to the reorganization rate characterized by the shift factor $a_T$ obtained from TTS (Figure 5G). For dissociative reorganization, hydrogel toughness decreases as reorganization dynamics increase (i.e. as $a_T$ decreases). This is expected, as lowering the activation energy of reorganization also lowers the energy barrier for rupture. In contrast, for associative reorganization, toughness is largely unaffected by the reorganization speed. Indeed, increasing the exchange rate via network composition does not affect the energy profile of the rupture pathway, hence, the energy required to break the network remains unchanged. These results highlight a fundamental distinction between associative and dissociative dynamics: while both govern relaxation, only dissociative mechanisms inherently trade off mechanical stregth for faster reorganization.

## 6. Conclusion

Our work reveals that associative crosslink exchange allows for the decoupling of network reorganization from thermal and mechanical stability in supramolecular hydrogels. This finding challenges the "strong means slow" paradigm for supramolecular networks, which holds only for dissociative reorganization. By controlling the mechanism of reorganization, specifically through associative crosslink exchange, it is possible to produce supramolecular networks that differ in reorganization speed by more than three orders of magnitude, yet have similar architecture, melting temperature, and rupture toughness. We provided a comprehensive description of the macroscale rheological behavior of both associative and dissociative DNA hydrogels, as well as a thermodynamic framework to rationalize this behavior. DNA is an ideal platform for this systematic comparison, as it enables the formation of both associative and dissociative hydrogels from the exact same DNA strands, only changing their proportions altered. Beyond DNA-based materials, these findings reveal a general design principle for supramolecular hydrogels: the reorganization mechanism is a key determinant of dynamic and mechanical properties. Associative crosslink exchange decouples reconfigurability from mechanical integrity, enabling the rational design of adaptive hydrogels that maintan thermal and mechanical stability. This work illustrates how reorganization pathways can be leveraged to control macroscale behavior across a wide range of dynamic materials. By identifying the reorganization mechanism as a key parameter, we aim to stirr the development of hydrogels that better mimic the mechanical and adaptive properties of biological tissues, while bridging the gap between self-healing ability and structural robustness.

## 7. Experimental Section/Methods

*Materials:*

ssDNA oligomers are purchased from Eurogentec, as summarized in Table 2. The enzymes T4 DNA ligase (2.5 WU·µL⁻¹), Exonuclease I (20 U·µL⁻¹), as well as deoxynucleotide triphosphates (dATP, dTTP, dGTP, and dCTP; 100 µM) and fluorescent deoxynucleotide triphosphates (Aminoallyl-dUTP-XX-ATTO-488 and Aminoallyl-dUTP-XX-ATTO-594) are purchased from Jena Bioscience. Φ29 DNA polymerase (10 U·µL⁻¹) is obtained from LGC Biosearch, and inorganic pyrophosphatase (0.1 U·µL⁻¹) from Thermo Fisher. Magnesium acetate ($MgAc_2$), sodium chloride (NaCl), calcium acetate ($CaAc_2$), silicone oil, tris(hydroxymethyl)aminomethane hydrochloride (Tris-HCl and Trizma buffer, pH 7.5), and poly-D-lysine (300 kg·mol⁻¹) are purchased from Sigma-Aldrich. Disodium ethylenediaminetetraacetate dihydrate (EDTA) is obtained from Euromedex. DNA strands are diluted and stored in TE buffer, which contains 10 mM Tris(hydroxymethyl)aminomethane (pH 8.0) and 1 mM EDTA. For hydrogel assembly and characterization, the TE buffer is supplemented with 100 mM NaCl and 10 mM magnesium acetate ($MgAc_2$), referred to as TENaMg buffer. DNA strands are stored at –25 °C in TE buffer.

Thermal ramps (up to 50 µL) are performed on a thermocycler (T100, Bio-Rad), while larger volumes are heated using a thermo shaker (PMHT, Grant Instruments). ssDNA concentrations are measured using a DeNovix DS-11 FX+ spectrophotometer, assuming a standard conversion factor of 33 µg·$OD_{260}^{-1}$. Capillary gel electrophoresis is conducted on a Qsep1 system (BiOptic Inc.), using standard DNA cartridges for short oligomers. Rheological measurements are carried out on an MCR 302 rheometer (Anton Paar) equipped with a sanded cone-plate geometry (diameter 8 mm, angle 3°, gap 50 µm; part number CP8-3) and a temperature-controlled oven (H-PTD 200). Fluorescence measurements are conducted using a Cary Eclipse fluorimeter (Agilent Technologies) equipped with a Peltier temperature controller. Fluorescence imaging is performed on an inverted microscope (Leica DMi8) using a 10× air objective (NA = 0.25), a LED excitation lamp (CoolLED pE-300), and a CMOS camera (C13440 ORCA Flash 4.0, Hamamatsu).

**Table2. Commercial DNA oligomer used, with their name, purification grade, and modifications.**

| | Name | Sequence 5’→3’ | Purification | Modification |
|---|---|---|---|---|
| **Linear Templates for RCA** | Tp_A[a] | GCTGTGCTCGGTGTTTTTTTTTTTTTTTTTTTTTTTTTTT TTTTGGTGCTCCTCGAC | HPLC | 5’-Phosphorylation |
| | $Tp_B_1$[a] | GTCGAGGTTTTTTTTTTTTTTTTTTTTTTTTTTTTTTTAC GTATAGGTCACCGAGCACAGC | HPLC | 5’-Phosphorylation |
| | $Tp_B_2$[a] | GTCGAGGAGCACCACTACTAACATTTTTTTTTTTTTT TTTTTTTTTTTTTTTGCACAGC | HPLC | 5’-Phosphorylation |
| **Ligation strands** | Lig_A | GCACAGCGTCGAGG | HPLC | None |
| | $Lig_B_{1/2}$ | CCTCGACGCTGTGC | HPLC | None |
| **RCA Primers** | Prim_A | GCACAGCGTCGA*G*G | RP-Cartridge - Gold | Phosphorothioated at position* |
| | $Prim_B_{1/2}$ | CCTCGACGCTGT*G*C | RP-Cartridge - Gold | Phosphorothioated at position* |
| **Fluorescent oligomer** | $Atto_{488}$ | TTTTTTTTTTTTTTTTTTTTT | HPLC | 5’ Atto488 |
| **Blocking Strands** | $Block\text{-}B_1$ | ACGTATAGGTCACCGA | SePOP Desalted | None |
| | $Block\text{-}B_2$ | AGCACCACTACTAACA | SePOP Desalted | None |
| **Molecular kinetics** | xA | CACCGAGCACAGCGTCGAGGAGCACC | HPLC | None |
| | $xB_1$ | CCTCGACGCTGTGCTCGGTGACCTATACGT | HPLC | None |
| | $xB_2$ | TGTTAGTAGTGGTGCTCCTCGACGCTGTGC | HPLC | None |
| | $xA_Q$ | CACCGAGCACAGCGTCGAGGAGCACC | HPLC | $3’\text{-}QXL_{570}$ |
| | $xB_{1F}$ | CCTCGACGCTGTGCTCGGTGACCTATACGT | HPLC | 5’-Atto565 |

*DNA synthesis:*

The DNA synthesis follows three main steps : the circularization of linear templates, the removal of unligated products, and template amplification via rolling circle amplification (RCA). A schematic representation of the DNA synthesis process, and the corresponding electropherogram are provided in Figure S1**.** For circularization, 4 µL of 10× T4 DNA Ligase buffer (Jena Bioscience; 500 mM Tris-HCl, pH 7.8 at 25 °C; 100 mM $MgCl_2$; 100 mM DTT; 10 mM ATP; 25 mg·mL$^{-1}$ BSA) is diluted in 32 µL of ultrapure water. Then, 2 µL of linear templates (100 µM in TE) and 2 µL of ligation strands (100 µM in TE) are added. The mixture is heated to 85 °C, then cooled to 25 °C at 0.5 °C·min$^{-1}$. After cooling, 2 µL of T4 DNA Ligase (2.5 WU·µL$^{-1}$) is added, and the mixture is incubated for 3 hours at room temperature. The enzyme is inactivated by heating to 70 °C for 20 minutes. Unligated products are removed using Exonuclease I (3 µL, 20 U·µL$^{-1}$) and Exonuclease III (1 µL, 200 U·µL$^{-1}$), which digest linear DNA. Circular templates are protected due to the absence of free ends. The mixture is incubated overnight at 37 °C, then heated to 80 °C for 40 minutes to inactivate the enzymes. Templates are purified using Amicon Ultra centrifugal filters (10 kDa cutoff, Merck Millipore) and rinsed three times with TE buffer. ssDNA concentrations are measured with a DS-11 FX+ spectrophotometer (DeNovix) and diluted to 1 µM. Capillary gel electrophoresis confirms circularization, as circular templates migrate faster than linear ones (Figure S1C). To verify enzyme activity, the same protocol is followed but T4 Ligase is replaced with water. If final DNA concentration falls below the spectrophotometer's detection limit, exonucleases are active and digest unligated DNA.

For RCA, 100 µL of 10× Φ29 DNA Polymerase buffer (Lucigen; 500 mM Tris-HCl, 100 mM $(NH_4)_2SO_4$, 40 mM DTT, 100 mM $MgCl_2$) is mixed with 770 µL ultrapure water, 50 µL circular templates (1 µM), 6 µL primer (10 µM), 20 µL Φ29 DNA Polymerase (10 U·µL$^{-1}$), and 1 µL pyrophosphatase (2 U·µL$^{-1}$). The mixture is incubated for 30 minutes at room temperature. Then, 10 µL of a custom dNTP mix (100 µM total, base ratios matched to product sequence) is added to initiate amplification. The reaction proceeds for 48 hours at 30 °C, followed by enzyme inactivation at 80 °C for 10 minutes. The product is concentrated using Amicon Ultra filters (30 kDa cutoff) and rinsed three times with 400 µL TE buffer. It is then rediluted in 200 µL TE, homogenized at 95 °C for 10 minutes, and quantified using the DS-11 FX+ spectrophotometer, assuming 33 µg·$OD_{260}^{-1}$ for ssDNA. The RCA stock solution is adjusted to 1 g·L$^{-1}$ in TE and stored at –20 °C. Capillary gel electrophoresis with standard S2 cartridges confirms the formation of long products (>5000 bp), though they cannot be accurately quantified (Figure S1B).

*Hydrogel assembly:*

A schematic representation of the hydrogel formation process, as well as relevant confocal fluorescence microscopy images, are provided in Figure S2. The hydrogel assembly process relies on liquid–liquid phase separation in the presence of divalent cations, here, calcium, to form dense all-DNA microgels.[61] The microgels are then washed to remove the divalent cations that induce phase separation and are subsequently concentrated by centrifugation. Above a DNA concentration of 10 g·L$^{-1}$, the microgels form a homogeneous hydrogel upon thermal treatment. In practice, RCA products (A, $B_1$, $B_2$), initially at 1 g·L$^{-1}$, are mixed in the desired proportions in TE buffer (10 mM Tris, pH 7.5, 1 mM EDTA) to reach a final concentration of 0.5 g·L$^{-1}$, typically in a 250 µL volume. The mixture is heated to 95 °C for 5 minutes in a thermal shaker with vigorous stirring for homogenization, then cooled at 4 °C for 5 minutes (Figure S2B). Once cooled, calcium acetate (0.5 M in ultrapure water) is added to reach a final $Ca^{2+}$ concentration of 50 mM. The solution is then heated again to 95 °C for 5 minutes, without agitation, to induce phase separation and form DNA microgels. After cooling to room temperature, the microgels are washed with TENaMg buffer (10 mM Tris, 1 mM EDTA, 100 mM NaCl, 10 mM $MgAc_2$) through three cycles of centrifugation (5000 g, 3 minutes) and redispersion in fresh buffer. The DNA concentration in the supernatant remains below 10 mg·L$^{-1}$, confirming efficient separation. Although TENaMg buffer does not induce phase separation at elevated temperatures, the DNA microgels remain stable due to supramolecular crosslinks between α/α* domains. Before the final centrifugation, 2 µL of the suspension is diluted in 18 µL of TE buffer and heated at 95 °C for 5 minutes to measure the DNA concentration. The microgel suspension is concentrated by removing supernatant until the target DNA concentration (2 wt%) is reached. The final suspension remains fluid and can be pipetted easily, for example to load into a rheometer. Finally, the hydrogel is formed in situ by heating the suspension to 95 °C for 5 minutes, dissolving the microgel particles (due to the absence of calcium), and cooling to room temperature to form a homogeneous 3D network.

Pristine RCA products are difficulet to handle due to their high viscosity even in the absence of crosslink. The treatment at 95°C in TE (10 min in the synthesis and 5 min in hydrogel assembly) help to reduce the the size of the RCA products which is necessary to facilitate handling. We show in Figure S2, the effect of the thermal treatment on the length of the RCA product and on ther resulting mechanical behavior of the DNA hydrogels.

*Rheology measurements:*

The inertia of the rheometer, with and without geometry, is recalibrated in air before each experiment. The surfaces of the rheometer in contact with the sample are thoroughly cleaned using ethanol and ultrapure water. To avoid slipping at high shear stress, the surfaces are coated with poly-D-lysine (300 kg·mol$^{-1}$) by loading in the rheometer 10 µL of poly-D-lysine solution at 1 g/L in ultrapure water, letting it adsorb for 5 min, rinsing three times with 20 µL of ultrapure water, and drying with compressed air. This protocol prevents interfacial slipping at high stress without affecting the bulk mechanical behavior of the hydrogels in the linear regime. We show in Figure S8 the effects of the surface treatment on the hydrogel's adhesion to the rheometer geometry and on startup shear measurements.

After surface treatment, the hydrogel is formed directly in the rheometer gap. About 10 µL of microgel suspension at the target concentration (2 wt%) is loaded directly onto the bottom plate of the rheometer and the target gap is imposed. Once the desired gap is reached (50 µm), any excess suspension is removed (less than 2 µL), and a few milliliters of low-viscosity silicone oil (47V100 from VWR, viscosity 100 cSt) are deposited at the sample rim to prevent evaporation. Finally, the oven (H-PTD200) is placed around the sample to ensure homogeneous heating of the sample. The hydrogel is formed in situ by heating the sample to 95 °C for 5 min and cooling it down to 25 °C at a rate of 0.05 °C/s. This heating step melts all supramolecular crosslinks and reforms the 3D network in a well-defined, reproducible state.

Measurements of storage and loss moduli (e.g., during temperature ramps) are performed in the linear regime by imposing oscillatory shear at 5% strain and a frequency $\omega = 1$ rad·s$^{-1}$. Frequency sweeps are performed at 5% strain for ω in the range of 0.1 to 10 rad·s$^{-1}$. Amplitude sweeps are conducted at $\omega = 1$ rad·s$^{-1}$ and between 0.1% and 1000% applied strain. Stress relaxation experiments are performed with a 5% step strain, recording the stress every 0.5 s for 7000 s. When imposing a temperature change, all ramps are conducted at 0.05 °C/s. Once the sample reaches the correct temperature, it is left to equilibrate for at least 5 min before measurements begin. Most experiments are conducted at 37 °C for two reasons: first, because it is a biologically relevant temperature for human-related research; second, because it accelerates reorganization and shortens the time required to measure relaxation and self-healing. Experiments conducted at 25 °C (room temperature) show similar behavior but exhibit slower relaxation dynamics.

*Time temperature superposition:*

For time temperature superposition tests, the hydrogel is first melted at 85 °C, before performing frequency sweeps at 85, 80, 75, 70, 65, 55, 45, 35, 25, and 15 °C. At each temperature, the sample is equilibrated for 5 min before starting the test. The the storage and loss moduli obtained at each temperature are shifted to form a master curve by multiplying frequency by a factor $a_T$ (horizontal shifting) and allowing for a modest vertical shift with a factor $b_T$. As an example, the full time-temperature supperposition process for a dissociative hydrogel (R=-0.5) is described in Figure S5 together with the curves obtained when $a_t$ is systematically maximized or minimized. We use this range in Figure 4B to quantify the errors on $a_T$. Note that we systematically present temperature-dependent properties obtained upon cooling rather than heating, in order to avoid history-dependent effects. Indeed, although the values of $a_t$ obtained upon heating and cooling are relatively close (Figure S5F), we always observe some small hysteresis, particularly near the transition temperature around 60 °C.

## 8. Acknowledgements

Wa thanks the ANR (Grant Number NR-20-CE06-0019 MIND) and the CNRS through the MITI interdisciplinary programs (Grant Number 249744) for the financial support. LC gratefully acknowledges support from the Institut Universitaire de France. RM thanks Marc Guerre and Gaëtan Bellot for the discussions.

## Data Availability Statement

The data that support the findings of this study are available from the corresponding author upon reasonable request.

**ToC :** How does reorganization impact the stability of supramolecular hydrogels? In contrast to dissociative crosslink exchange, implementing associative exchange in macroscale DNA hydrogels enables decoupling of reorganization dynamics from thermal and mechanical stability. This establishes the reorganization mechanism, and its underlying thermodynamic pathway, as a key design parameter for creating robust, reconfigurable soft materials.

**Decoupling dynamics and crosslink stability in supramolecular hydrogels using associative exchange.**

**ToC Figure :**

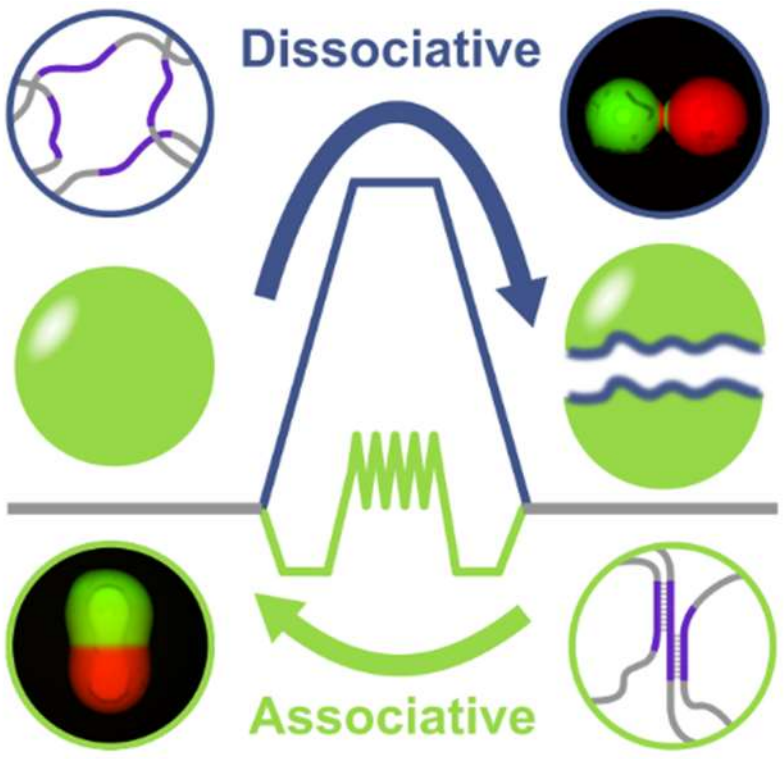

# Supporting Information

**Decoupling dynamics and crossink stability in supramolecular hydrogels using associative exchange.**

*Pierre Le Bourdonnec, Charafeddine Ferkous, Leo Communal, Luca Cipelletti, and Rémi Merindol*.*

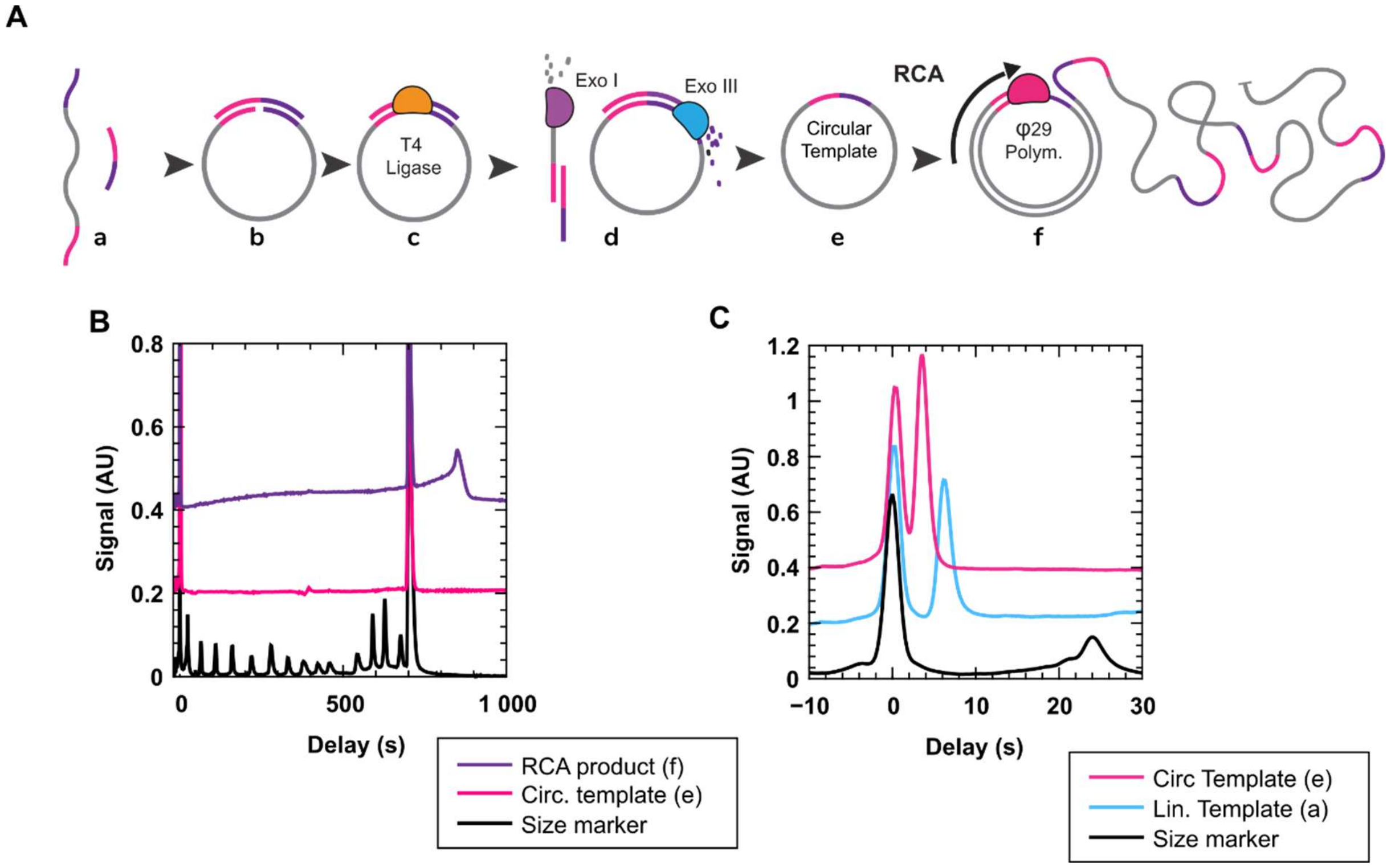

**Supporting Figure S1: DNA synthesis.** A) Schematic representation of the steps involved in the rolling circle amplification process. B) Representative electropherogram obtained via capillary gel electrophoresis, showing the circularized template (pink), the RCA product (purple), and the commercial DNA ladder (black) (From Bioptic Inc., Ref.109300, fragment sizes : 50, 100, 150, 200, 250, 300, 400, 430, 450, 500, 750, 1100, 1800 , and 3000 base pairs). The two peaks at 0 and 700s correspond to the alignment markers at 20 and 5000 base pairs, respectively. Data have been vertically offset for clarity. Due to the large time scale, the peak corresponding to the circular template is too close to the alignment marker to be distinguished. The RCA product, in contrast, displays a broad peak centered around 800s, beyond the 5000 base pair alignment marker. C) Zoomed view of the first 30s of the capillary electrophoresis signal, highlighting the peak of the circularized template (pink, at 4s) which appears slightly before the linear template peak (blue, at 8s).

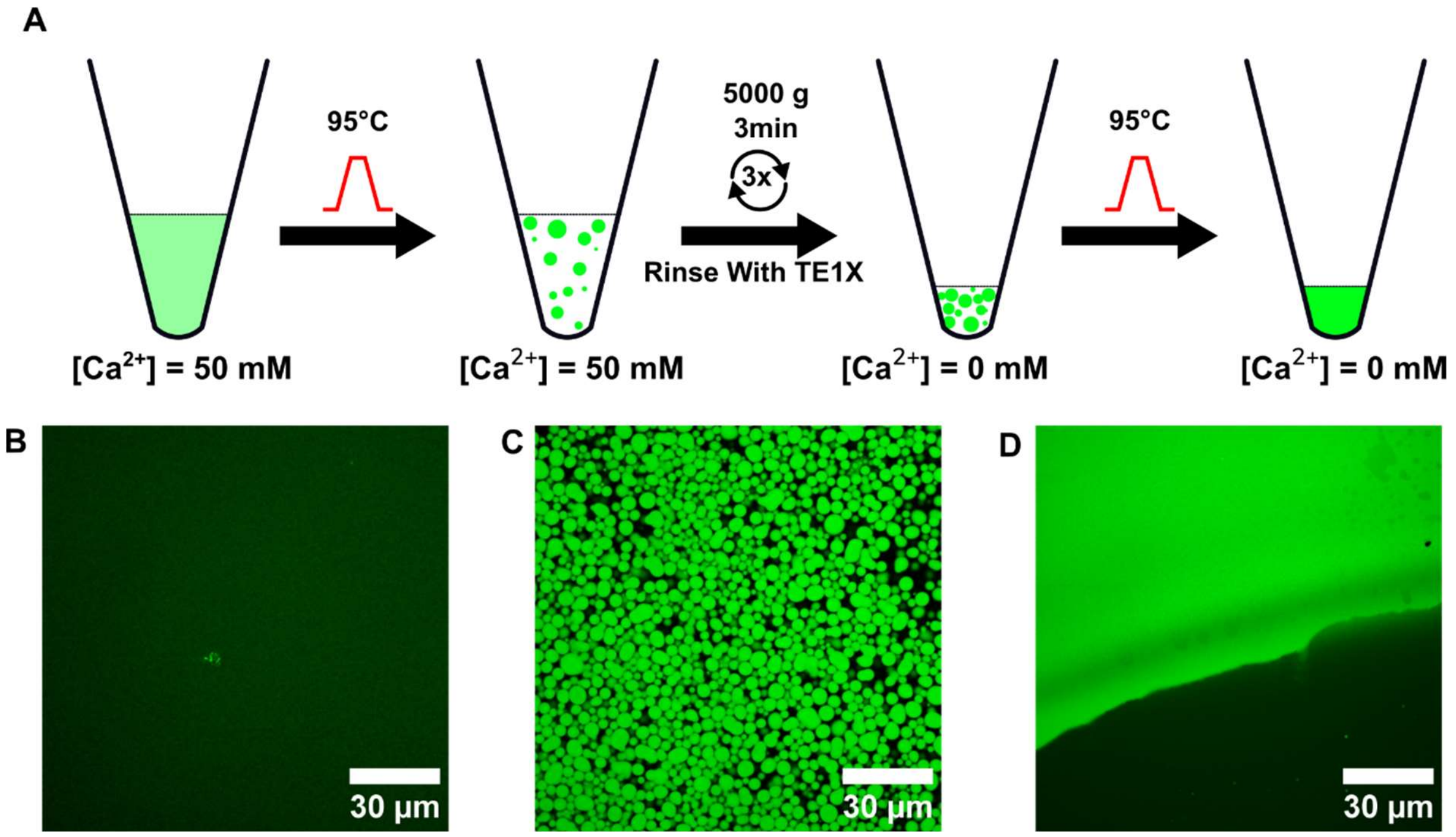

**Supporting Figure S2: Hydrogel assembly.** A) Schematic representation of the steps involved in the hydrogel assembly process. B–D) Confocal microscopy images of an $Atto_{488}$-labeled RCA product: as synthesized (B), after phase separation (C), and after hydrogel formation (D).

**Supporting Note S1: Preparation of fluorescent DNA hydrogels.** The synthesis of fluorescent RCA products follows the protocol described in the main text, with the addition of 2 µL of fluorescent dNTP (10 mM; Aminoallyl-dUTP-XX-ATTO-488 or Aminoallyl-dUTP-XX-ATTO-594) to the dNTP mix. The hydrogel assembly protocol remains unchanged, except that 30% of one RCA product (typically sequence A) is replaced by a fluorescently labeled counterpart of the same sequence prior to microgel formation. For hydrogel melding experiments, spherical hydrogels are formed by introducing 3 µL of DNA microgel suspension (20 $g \cdot L^{-1}$) into mineral oil preheated at 95 °C, incubating for 3 minutes, and then cooling to room temperature. After cooling, the spherical microgels can be manipulated using tweezers.

The microgels are manually inserted into a 96-well glass-bottom plate filled with fresh mineral oil, and pairs of microgels are gently brought into contact using tweezers. The plate is stored in a thermostated oven at 30 °C for 3 days. Imaging is performed regularly at room temperature using a Leica DM8 wide-field fluorescence microscope with GFP and RFP filter sets.

**Supporting Note S2: Length of the RCA products**

We reported in a previous article that DNA degrades at high temperature.[61] Here, we take advantage of this process to control the size of the RCA products. Reducing the size of the RCA products is necessary to facilitate handling by lowering the viscosity of the strand suspensions, although it also impacts the resulting physical properties. Using capillary gel electrophoresis, we confirm that heating the RCA product in TE buffer at 95 °C decreases its molecular weight (Figure S3A). Initially, the product is so viscous that it does not migrate in the gel and shows no significant signal. After 5 minutes at 95 °C, a broad peak appears and progressively shifts to shorter migration times as the heating duration increases, indicating the formation of shorter DNA strands. This behavior clearly signals the thermal shortening of the RCA product. After 30 minutes at 95 °C, oscillations emerge in the electropherogram, corresponding to the repeat units of the RCA product. These oscillations suggest that cleavage preferentially occurs at specific sites within the sequence.

As we aim to form hydrogels from long DNA strands rather than short oligonucleotides, we focus on short heating durations between 0 and 30 minutes at 95 °C. The loss and storage moduli of these hydrogels, measured as a function of heating time, are displayed in Figure S3B. The elastic modulus systematically decreases with increasing heating time, while the loss modulus remains nearly constant and only begins to decrease after 20 minutes of heating. This trend is expected, as shorter DNA strands form fewer entanglements and possess more free dangling ends that do not contribute to the elasticity of the network. Given the high polydispersity of the RCA product under all conditions, we do not explore this effect further. In all experiments, the RCA product undergoes heating for approximately 20 minutes at 95 °C, which ensures both ease of handling and reproducible mechanical behavior.

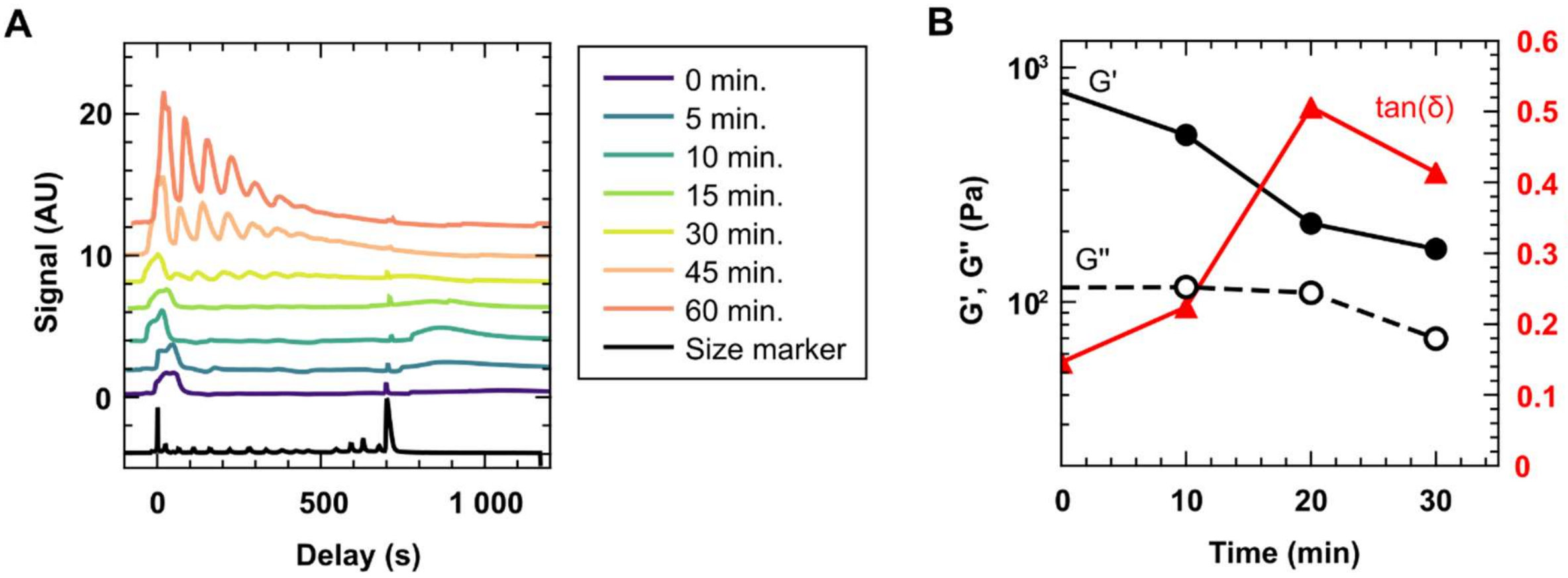

**Supporting Figure S3:** A) Electropherogram of RCA product A after 0 to 60 minutes in TE buffer at 95 °C. The size of the RCA product progressively decreases with heating time. The alignment markers are 20 and 5000 base pairs, and the size marker (Bioptic Inc., Ref. 109300) includes fragments of 50, 100, 150, 200, 250, 300, 400, 430, 450, 500, 750, 1100, 1800, and 3000 base pairs. B) Effect of heating time on the linear mechanical behavior of an R = -1 hydrogel at 1.5 wt%.

**Supporting Note S4: Fitting stress relaxation**

Stress relaxation experiments are fitted with a stretched exponential decay, Equation 2 of the main text that we recall here for simplicity:

$$\sigma(t) = \gamma G(t) + s = \gamma G_0 \exp\left[-\left(\frac{t}{\tau_r}\right)^{\beta}\right] + s, \quad \text{Supporting Equation S1}$$

where the fitting parameters are $G_0$, $\tau_r$, $\beta$, and $s$. While in principle $s = 0$, in some cases we observed a small and constant remaining stress (<2 Pa) at long timescale. Its effect is negligible on static hydrogels (R≤0) as we do not observe the full relaxation. However, for dynamic hydrogels, this residual stress systematically leads to an underestimation of the stretching exponent β. We assumed this remaining stress to be an artifact and allowed for a small baseline correction ($s$ <2 Pa) in the fits. Note that in Figure 3C of the main text we show the raw $G(t)/G_0$ data without baseline correction, from which one can appreciate that such correction is indeed very small.

From the fitted $G(t)$, the integral relaxation time is calculated as

$$\tau_i(G_0, \tau_r, \beta) = \frac{\int_0^\infty G(t)\mathrm{d}t}{G_0} = A\int_0^\infty \exp\left[-\left(\frac{t}{\tau_r}\right)^{\beta}\right]\mathrm{d}t \; = \; A\tau_r\frac{\Gamma(1/\beta)}{\beta}$$

Supporting Equation. S2

where $\Gamma(x)$ is the Gamma function that generalizes the factorial function to non-integers arguments. Note that in writing Equation S2 we have explicitly included the dependence on $G_0$: while $A = 1$ by definition (compare Equation S2 to Equation S1), this prefactor is affected by an uncertainty $\delta A = \delta G_0/G_0$, due to the error $\delta G_0$ on the fitting parameter $G_0$. The error bars on $\tau_i$ shown in the main text correspond to the propagation of the errors on $G_0$, $\tau_r$, $\beta$, obtained from the fits. For a set of fitting parameters $\bar{G}_0, \bar{\tau}_r, \bar{\beta}$, one has (Equation S3) :

$$\delta\tau_i = \sqrt{\left[\frac{\partial\tau_i}{\partial G_0}\right]^2_{\bar{G}_0,\bar{\tau}_r,\bar{\beta}}\delta^2 G_0 + \left[\frac{\partial\tau_i}{\partial\tau_r}\right]^2_{\bar{G}_0,\bar{\tau}_r,\bar{\beta}}\delta^2\tau_r + \left[\frac{\partial\tau_i}{\partial\beta}\right]^2_{\bar{G}_0,\bar{\tau}_r,\bar{\beta}}\delta^2\beta}$$

$$= \sqrt{\frac{\delta^2 G_0}{G_0^2} + \left[\frac{\Gamma(1/\beta)}{\beta}\right]^2_{\bar{\beta}}\delta^2\tau_r + \left[\frac{\partial\tau_i}{\partial\beta}\right]^2_{\bar{G}_0,\bar{\tau}_r,\bar{\beta}}\delta^2\beta}$$

Supporting Equation. S3

where the derivative of $\tau_i$ with respect to $\beta$ is calculated numerically.

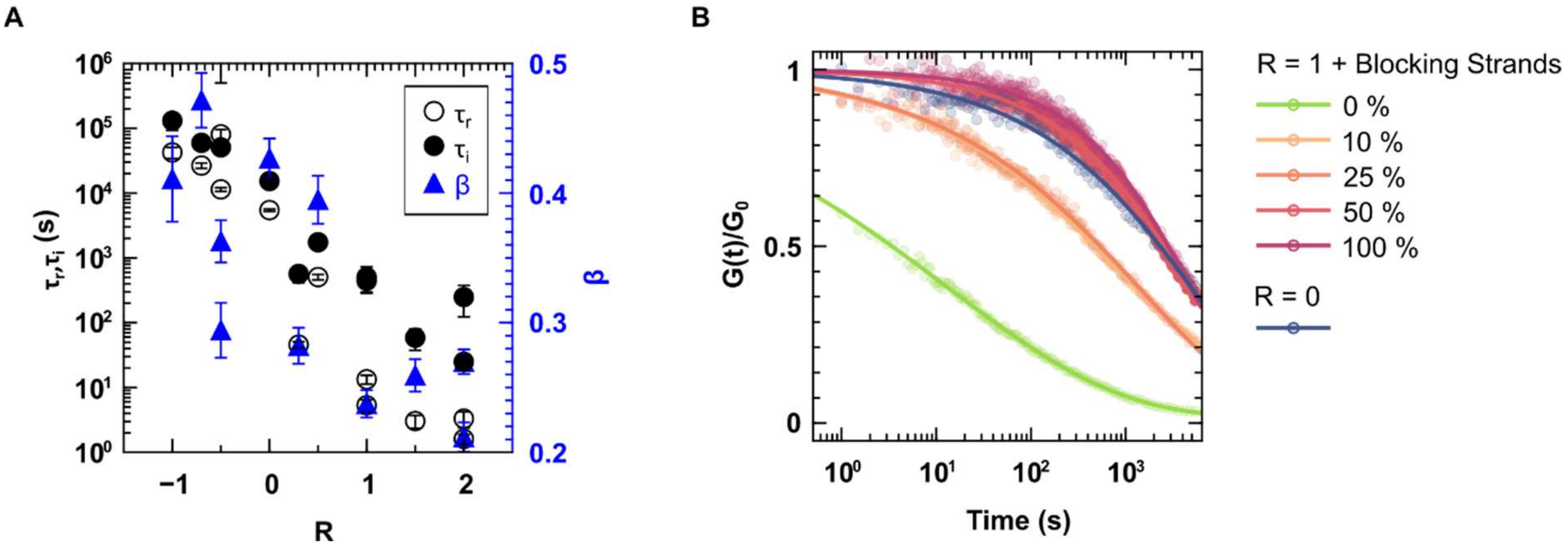

**Supporting Figure S4:** A) Values of the *1/e* relaxation time ($\tau_r$), integral relaxation time ($\tau_i$) and stretching exponent ($\beta$) obtained from fitting the stress relaxation data using a stretched exponential decay. B) Stress relaxation experiments for associative hydrogels ($R$=1) with increasing proportion of blocking strands (5% strain, 37 °C) and for dissociative hydrogels ($R$=0). Symbols represent experimental stress measurements; solid lines correspond to stretched exponential fits.

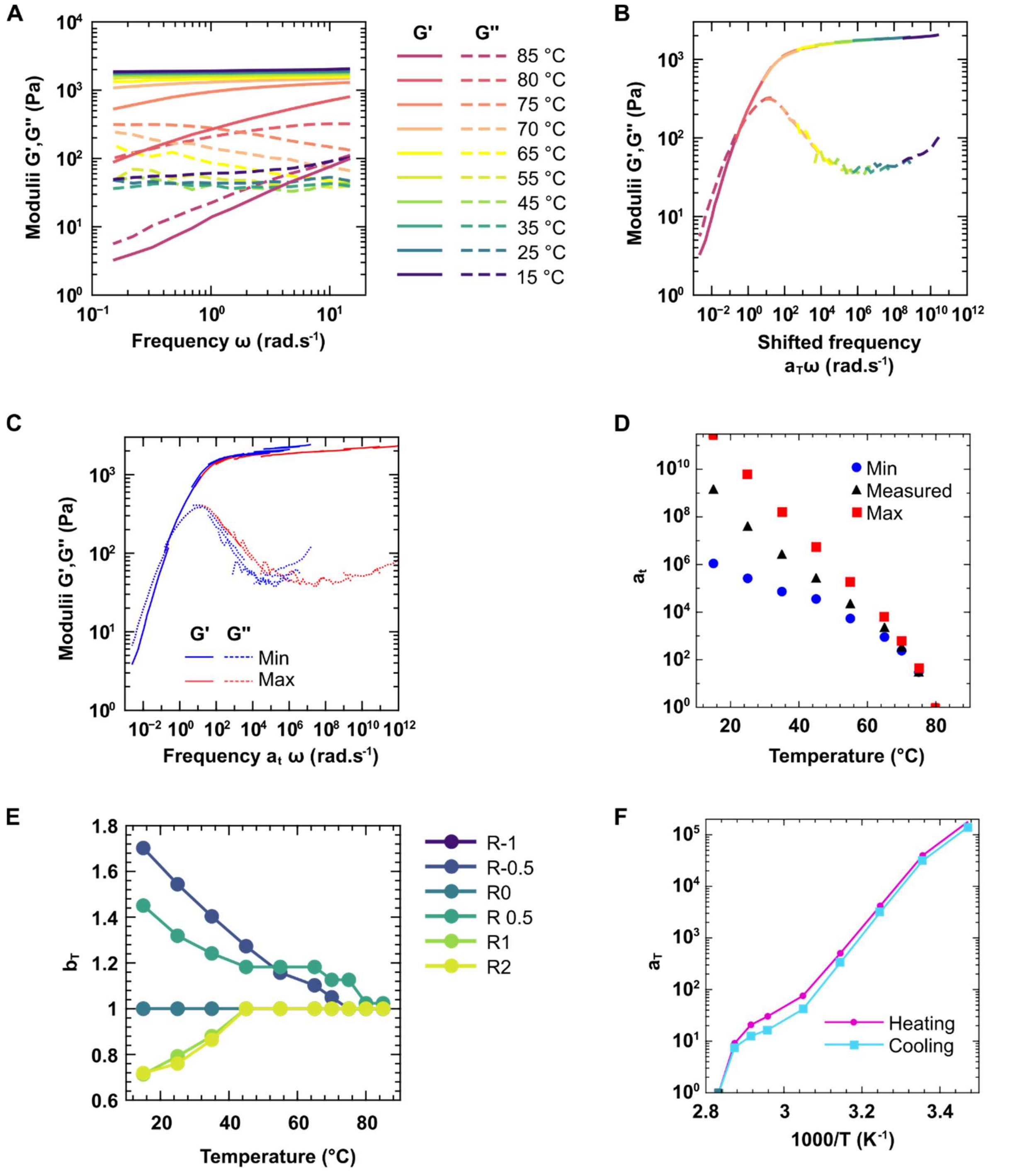

**Supporting Figure S5: Time Temperature Superposition** A) Evolution of the storage modulus ($G'$) and loss modulus ($G''$) obtained from frequency sweeps between 0.1 and 10 rad·s$^{-1}$ for a static hydrogel ($R$ = -0.5) at temperatures ranging from 85 °C to 15 °C. B) Corresponding master curve obtained after manual alignment of all frequency sweeps. C) Examples of master curves that are overshifted (red) or undershifted (blue). Note that at low temperatures, the moduli are relatively insensitive to changes in both frequency and temperature, which complicates accurate superposition. D) Corresponding shift factor $a_t$ measured for the actual shift (values reported in the main text), as well as for overshifted and undershifted curves. As expected, a broad range of $a_t$ can be used to superimpose the curves of such a static hydrogel at low temperature. E) Values of the vertical shift $b_t$ applied to obtain the master curves. F) Examples of shift factors obtained either by heating (pink line) or by cooling (blue line) the sample.

**Supporting Note S5: Molecular kinetics by fluorometry**

**Experimental design:** We aim to model the associative crosslink exchange and monitor the strand displacement reaction at the molecular scale. To achieve this, we construct a model strand displacement reaction using commercially synthesized DNA oligomers, tracked quantitatively via fluorescence labeling. The model reaction, shown in Figure S6A, consists of two parts: a background strand displacement reaction, which mimics associative crosslink exchange, and a reporter reaction, which enables fluorescence-based monitoring of this exchange. To avoid confusion with the RCA products used in hydrogel assembly, all short commercial oligomers used in this experiment are labeled with the prefix "x". In the background reaction, the oligomers xA, $xB_1$, and $xB_2$ share the same sequences as the RCA products A, $B_1$, and $B_2$, respectively, with the domains α, β1, $\beta_2$ on A; α*, $\beta_1$* on $B_1$; and $\beta_2$*, α* on $B_2$. However, they lack the $A_{20}$ spacer domain and consist of a single repeat unit. These sequences can be mixed at the same stoichiometric ratio *R* as in the RCA-based system, but they do not form a percolating 3D network. The reporter reaction involves two modified oligomers: $xA_Q$, which has the same sequence as xA and is functionalized with a quencher ($QXL_{570}$) at its 3′-end; and $xB_{1F}$, which has the same sequence as $xB_1$ and is labeled with a fluorophore ($Atto_{565}$) at its 5′-end. When $xA_Q$ and $xB_{1F}$ hybridize, the fluorophore and quencher are brought into close proximity. In this configuration, the complex is non-fluorescent due to Förster Resonance Energy Transfer (FRET), whereby the fluorophore de-excites via energy transfer rather than photon emission. If $xB_{1F}$ is displaced by $xB_2$, the fluorophore becomes spatially separated from the quencher, FRET efficiency drops, and fluorescence increases. Because of the high efficiency of FRET, this system functions as a binary ON/OFF fluorescence sensor for the $xA_Q/xB_{1F}$ duplex. The fluorescence increase is therefore proportional to the loss of the $xA_Q/xB_{1F}$ complex, and to the formation of free $xB_1$ and new duplexes $xA/xB_{1F}$, both of which are fluorescent. Assuming that the duplexes $xA_Q/xB_{1F}$, $xA/xB_1$, and $xA/xB_2$ share similar reaction kinetics, and that the addition of a small amount of reporter does not perturb the global equilibrium, we can quantify the exchange kinetics of the background reaction by monitoring the fluorescence increase immediately after adding a small amount of reporter complex.

**Protocol.** The background reaction is prepared by mixing oligomers xA, $xB_1$, and $xB_2$ at the target stoichiometric ratio *R*, maintaining a constant total concentration of duplexes (xA/xB) at 2 μM in 400 μL of TENaMg buffer (10 mM Tris, 1 mM EDTA, 100 mM NaCl, 10 mM $MgAc_2$).

Separately, we prepare the fluorescent reporter by mixing $xA_Q$ and $xB_{1F}$ in a 1:1 molar ratio to form a duplex at 0.5 µM final concentration. Both the background and reporter mixtures are heated at 85 °C for 5 minutes, then cooled and equilibrated at the target temperature (between 15 °C and 55 °C) before starting the reaction. To initiate the fluorescence assay, we first add 8 µL of the reporter duplex to a 30 µL fluorescence cuvette (Hellma Analytics), which is then placed in a temperature-controlled fluorimeter (Cary Eclipse, Agilent, equipped with a Peltier temperature controller). We begin fluorescence acquisition with excitation at 565 nm, emission at 590 nm, and a recording interval of 0.1 s. After a few seconds, we add 200 µL of the thermalized background mixture into the cuvette and mix rapidly by pipetting the solution up and down five times. Fluorescence is then recorded for 3000 s at constant temperature. We determine the maximum theoretical fluorescence ($F_{max}$) independently by measuring the fluorescence of a similar mixture (reporter + background) after incubation at 85 °C for 5 minutes, ensuring full equilibration. This measurement is performed at 25 °C.

For experiments involving long RCA products (reported in Figure 4C of the main text), we replace the background oligomers (xA, $xB_1$, $xB_2$) with the corresponding RCA products (A, $B_1$, $B_2$) at the same molar concentration of repeat sequence.

**Data treatment.** Typical fluorescence time traces (normalized by $F_{max}$) are shown in Figure S6B as a function of the stoichiometric ratio *R*, and in Figure S6C as a function of temperature. The reaction is designed to follow first-order kinetics, since the concentration of displacing strand $xB_2$ is much higher than that of the reporter complex $xA_Q/xB_{1F}$. Therefore, we expect fluorescence *F(t)* to follow:

$$F(t) = F_{max} * (1 - e^{-\frac{t}{k_{app}}}) \qquad \text{Supporting Equation S4}$$

where *t* is time, and $k_{app}$ is the apparent exchange rate, which depends on the concentration of available $xB_2$ (thus on the stoichiometric ratio *R* of the background reaction). We extract $k_{app}$ by plotting $ln(F_{max}-F(t))$ and fitting the initial slope of the curve (Figure S6D). Deviations from linearity at longer times are expected, as $F_{max}$ is determined independently and may vary due to sequence impurities or pipetting errors. This first-order kinetic assumption holds only if strand displacement is the sole mechanism of $xA_Q/xB_{1F}$ dissociation. At high temperatures, where DNA duplexes can thermally melt, this assumption breaks down. To validate the temperature range, we use an $R = 0$ background reaction and monitor fluorescence as a function of temperature. As shown in Figure S6E, the normalized fluorescence increase remains below 10% up to 55 °C. We therefore set 55 °C as the upper limit for our fluorescence experiments.

**Blocking strand.** We confirm in Figure S6F that blocking strands can inhibit the associative strand exchange. In this control, we compare an $R = 1$ background reaction with and without blocking strands. For the blocking condition, we add 22 μL of a mixture containing 20 μM each of Block-$B_1$ and Block-$B_2$ (in TENaMg buffer) 100 seconds after the start of the experiment. This results in a final blocking strand concentration of approximately 4 μM, corresponding to 100% saturation of available binding sites. Upon addition, the fluorescence increase slows down dramatically, indicating that the blocking strands efficiently inhibit strand exchange. In contrast, in the control reaction without blockers, fluorescence continues to increase steadily.

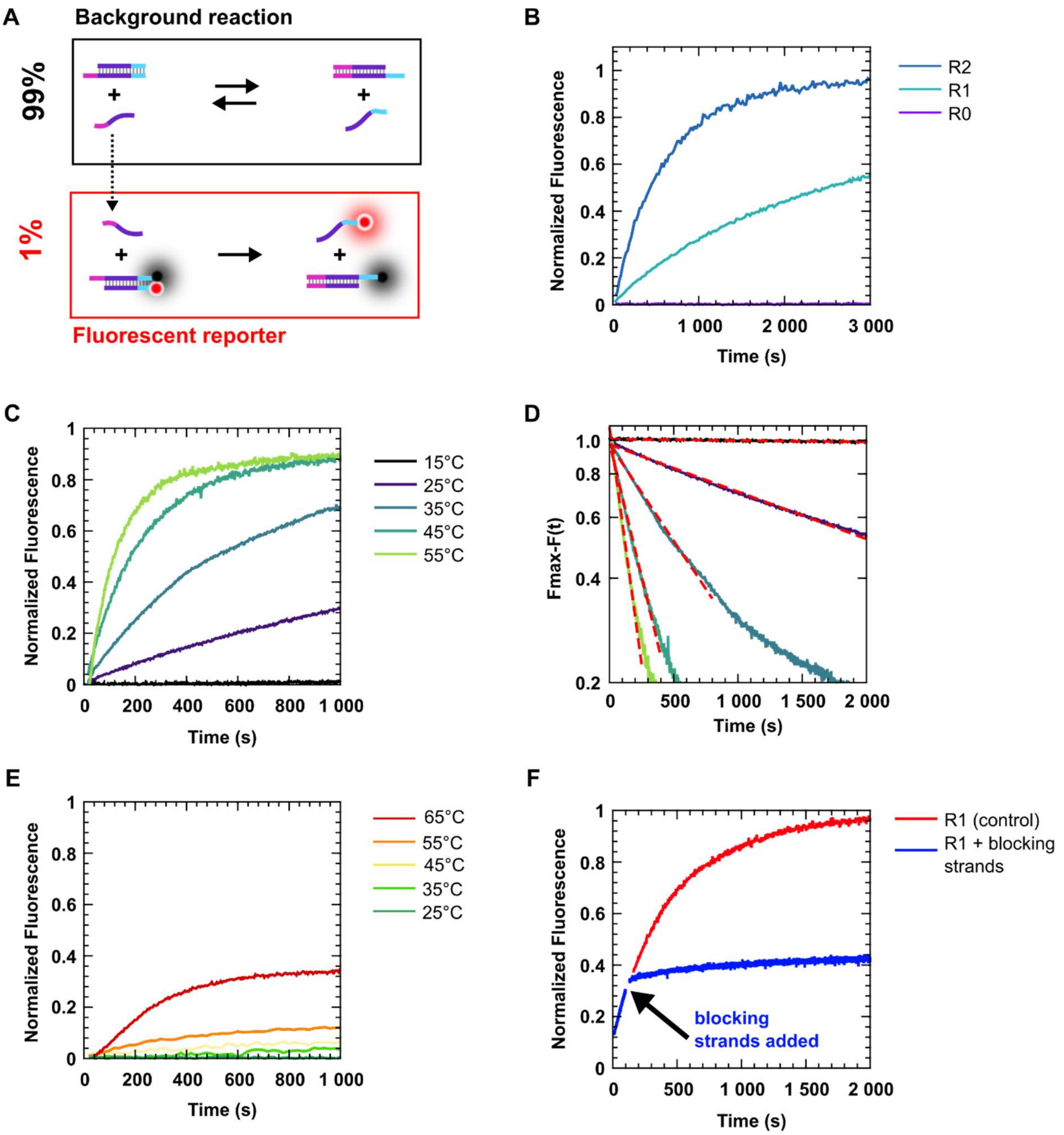

**Supporting Figure S6:** (A) Schematic representation of the experimental design used to monitor strand displacement via fluorescence. The black box shows the background reaction, which consists of a mixture of unlabeled DNA oligomers (xA, $xB_1$, $xB_2$) mimicking the crosslinking domains of RCA products A, $B_1$, and $B_2$ (same sequences but without the $A_{30}$ spacer). The red box shows the fluorescent reporter, a duplex formed by $xA_Q$ (quencher-labeled) and $xB_{1F}$ (fluorophore-labeled). Upon displacement of $xB_{1F}$ by $xB_2$, the fluorophore separates from the quencher, stopping FRET and resulting in an increase in fluorescence. B) Evolution of fluorescence with time after adding the fluorescent reporter at $t = 0$, measured at 25 °C, for different stoichiometric ratios $R$ of the background reaction. C) Evolution of fluorescence with time after adding the fluorescent reporter at $t = 0$, measured at different of temperatures for a constant stoichiometric ratio $R = 1$ for the background reaction. D) Example of the linear fits (dashed red lines) of the plots $ln(F_{max}-F(t))$ versus $t$, used to extract the apparent exchange rate $k_{app}$, assuming first-order kinetics. The data correspond to the same experiments and color code as in (B). E) Control experiments showing a negligible fluorescence increase below 55 °C for a background reaction stoichiometric ratio $R = 0$, confirming that the strand displacement reaction is negligible under these conditions. F) Validation of strand exchange inhibition by blocking strands. Upon addition of 100% blocking strands at t = 100 s in a reaction with $R = 1$, the fluorescence increase slows dramatically, confirming efficient suppression of the exchange. In contrast, the control without blockers continues to exhibit fluorescence growth.

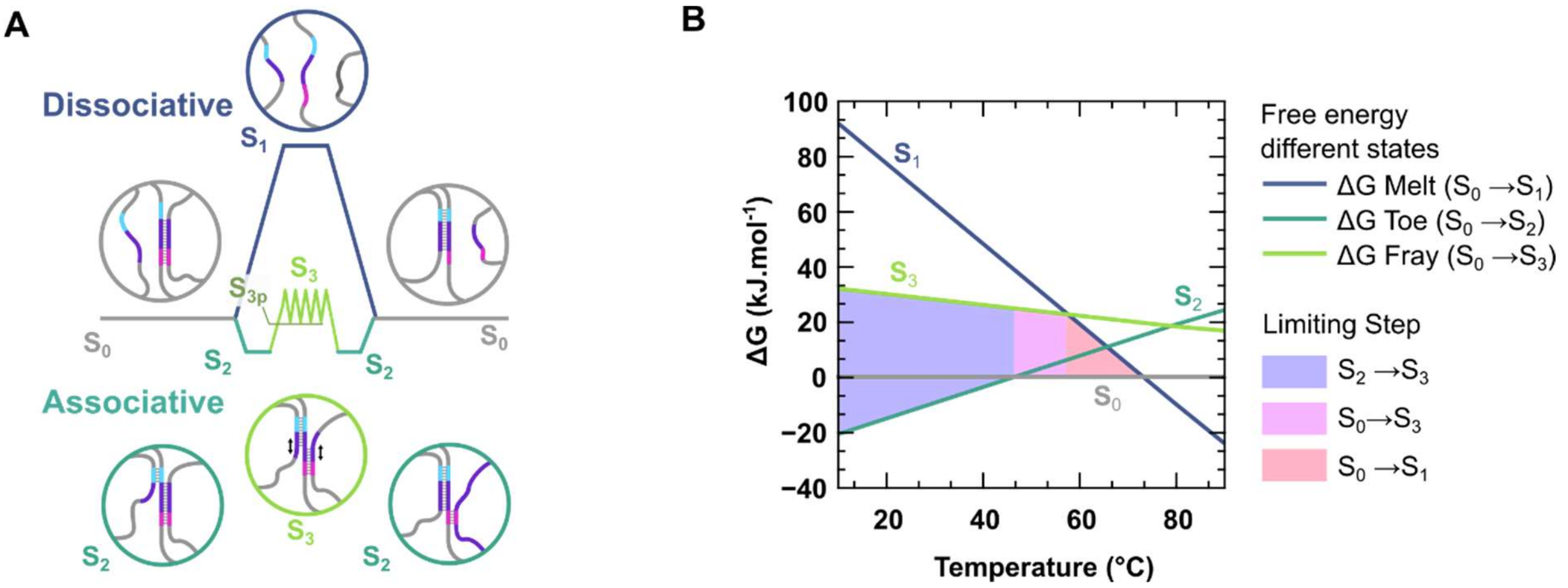

**Supporting Figure S7: Thermodynamic profiles** A) Schematic thermodynamic profile of dissociative (blue) and associative (green) crosslink exchange pathways, including the names and molecular representations of the states. B) Temperature-dependent evolution of the free energy levels ($\Delta G$) for the various assembly states. Values are calculated relative to $S_0$ using theoretical thermodynamic parameters for oligomers in solution (see Table S1). The shaded regions indicate the rate-limiting steps for crosslink exchange in an associative hydrogel ($R > 0$). Three colors highlight distinct regimes: at low temperature (blue), the limiting step is strand invasion following toehold hybridization ($S_2 \rightarrow S_3$); at intermediate temperatures (pink), it is direct invasion without toehold hybridization ($S_0 \rightarrow S_3$); and at high temperature (red), it is the melting of the duplex during dissociative exchange ($S_0 \rightarrow S_1$).

**Supporting Note S6: Theoretical thermodynamic profiles**. The thermodynamic properties of DNA hybridization are well established.[59] Here, we employed the UNAFold platform (https://www.unafold.org) to compute the theoretical thermodynamic parameters for hybridization of the various DNA domains.[52,62] For these simulations, we set the NaCl concentration to 100 mM and the concentration of each DNA oligomer to 400 µM, corresponding to the concentration of repeat units in a 2 wt% hydrogels. Since UNAFold does not support strand displacement reactions, we incorporated additional thermodynamic parameters from literature. The free energy profile of strand displacement has been extensively described by Winfree and coworkers in their study on strand displacement kinetics.[56] They reported a total free energy change of 30 kJ·mol$^{-1}$, decomposed into two components: (1) an initiation energy barrier, associated with the formation of a three-stranded junction, corresponding to a plateau height of $\Delta G_p$ = 8 kJ·mol$^{-1}$, and (2) an energy cost for progressive base-pair unzipping during junction migration, corresponding to a sawtooth pattern of $\Delta G_s$ = 22 kJ·mol$^{-1}$, as illustrated in Figure S7A (green line, states S3p and S3 respectively). To model the temperature dependence of the energy profile (see Figure 4E, in the main text), we further decomposed the free energy ($\Delta G$) into enthalpic ($\Delta H$) and entropic ($\Delta S$) contributions. For the plateau, we used Winfree's estimated values from their Supplementary Information (page 13): $\Delta H_p$ = 54 kJ·mol$^{-1}$ and $\Delta S_p$ = 154 J·mol$^{-1}$·K$^{-1}$.[56] For the sawtooth component, no direct values for $\Delta H$ and $\Delta S$ are available in the literature. To estimate these parameters, we referred to the unified nearest-neighbor thermodynamic model by SantaLucia,[59] choosing GA/CT base pairs as a representative case with intermediate hybridization strength. For these, the free energy of hybridization is $\Delta G_{GA/CT} \approx$ 6 kJ·mol$^{-1}$. Notably, this value is significantly lower than the 22 kJ·mol$^{-1}$ calculated by Winfree for strand displacement. We hypothesize that this discrepancy arises from a reduced entropic gain during strand migration compared to DNA melting. In DNA melting, base-pair disruption results in strand separation, increasing system entropy via a greater number of accessible microstates. In contrast, during branch migration, base unstacking occurs between already hybridized duplexes, contributing little to the entropy change. Accordingly, we propose the following decomposition for the sawtooth free energy: $\Delta H_s$ = 34 kJ·mol$^{-1}$ = $\Delta H_{S3p \rightarrow S3}$, taken directly from SantaLucia,[59] and $\Delta S_s$ = 39 J·mol$^{-1}$·K$^{-1}$= $\Delta S_{S3p \rightarrow S3}$, adjusted to yield $\Delta G_s$ = 22 kJ·mol$^{-1}$ at 25 °C, under the assumption that branch migration proceeds via single base-pair rupture. The resulting thermodynamic data, used to construct the free energy diagrams in Figure 4D and Figure S7, are summarized in Table S1. The good agreement between the theoretical enthalpy variation and the activation energy derived from fluorometric measurements (Figure 4C) further supports this hypothesis.

**Supporting Table S1 : Thermodynamic Data.**

| | ΔH (kJ.mol$^{-1}$) | ΔS (J.mol$^{-1}$.K$^{-1}$) | ΔG at 25°C (kJ.mol$^{-1}$) | Reference |
|---|---|---|---|---|
| **$S_0$ (Reference)** | 0 | 0 | 0 | UNAFold, [Oligo] = 400 µM , [$Na^+$] = 100 mM. |
| **$S_0 \rightarrow S_1$** | +735 | +2068 | +119 | UNAFold, [Oligo] = 400 µM , [$Na^+$] = 100 mM. |
| **$S_0 \rightarrow S_2$** | -176 | -552 | -12 | UNAFold, [Oligo] = 400 µM , [$Na^+$] = 100 mM. |
| **$S_2 \rightarrow S_{3p}$ (plateau height)** | +54 | +155 | +8 | [56] |
| **$S_{3p} \rightarrow S_3$ (sawtooth amplitude)** | +34 | +40 | +22 | [56,59] |
| **$S_2 \rightarrow S_3$** | +88 | +195 | +30 | $S_2 \rightarrow S_{3p}$ + $S_{3p} \rightarrow S_3$ |
| **$S_0 \rightarrow S_3$** | -88 | -357 | +18 | $S_0 \rightarrow S_2$ + $S_2 \rightarrow S_3$ |

**Supporting Note S7: Poly-Lysine treatment and startup shear experiments**

The effect of poly-D-lysine treatment is most evident in start-up shear experiments (Figure S8). Without treatment, the hydrogel breaks below 5 kPa (Figure S8B), while an hydrogel from the same batch, measured after treating the rheometer tools with poly-D-lysine, withstands stresses approaching 40 kPa. In the linear regime (below 100% strain), both curves overlap, confirming that poly-D-lysine treatment does not affect the bulk mechanical properties of the hydrogels. We also use fluorescently labeled DNA hydrogels and a UV lamp (365 nm, 6 W) to visualize the sample after mechanical rupture. Without poly-D-lysine treatment, the sample remains in one piece on the bottom plate of the rheometer, indicating detachment from the upper tool. With poly-D-lysine treatment, the sample fractures in the bulk into multiple pieces, with portions adhering to both rheometer surfaces after failure (Figure S8A). This result confirms that strong adhesion to the rheometer tools is necessary to measure the true rupture strength of DNA hydrogels. Start-up shear experiments are performed at a high strain rate ($1\ s^{-1}$), which minimizes the effect of network reorganization on rupture behavior. After each test, the hydrogel is heated at 85 °C for 5 minutes to reform the network. The first rupture typically results in a higher ultimate stress than subsequent ruptures (Figure S8C), which yield consistent values. Therefore, we only report start-up shear data obtained after the first rupture.

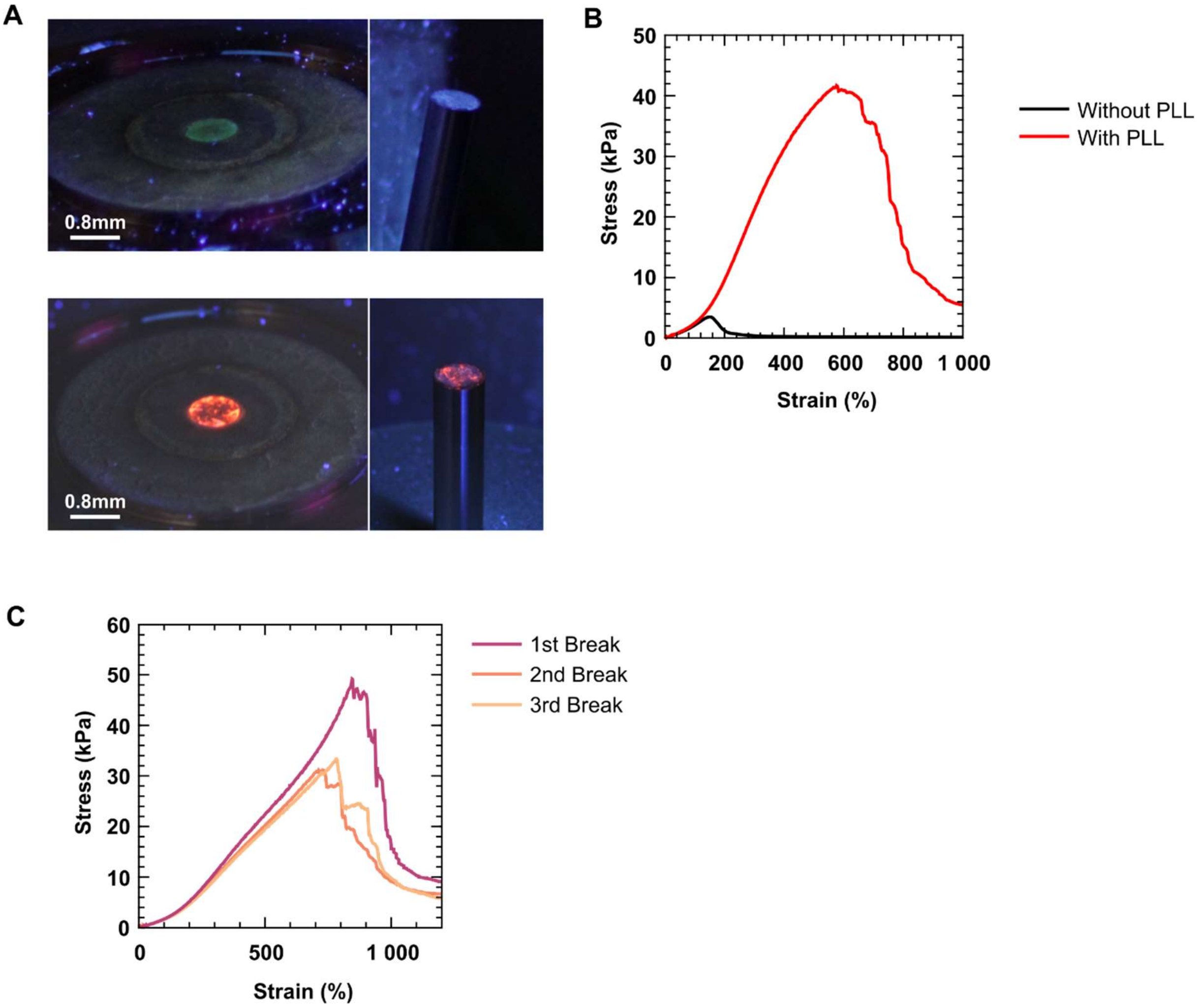

**Supporting Figure S8:** A) Photographs of fluorescent DNA hydrogels under UV illumination (365 nm, 6 W) after start-up shear experiments, shown without (top) and with (bottom) poly-D-lysine treatment. B) Effect of poly-D-lysine treatment on start-up shear experiments performed on an $R$ = -1 hydrogel at a strain rate of 1 $s^{-1}$. C) Multiple start-up shear experiments performed on the same $R$ = -1 hydrogel, with the network reformed at 85 °C between tests.